\documentclass{article}[12pt]
\pdfoutput=1
\usepackage{jheppub}
\usepackage[T1]{fontenc}
\usepackage[utf8]{inputenc}
\usepackage[english]{babel}
\usepackage[babel]{csquotes}
\usepackage{epigraph}
\usepackage{amsfonts}
\usepackage{mathrsfs}
\usepackage{microtype}
\usepackage{booktabs}
\usepackage{array, tabularx}
\usepackage{chngpage}
\usepackage{braket}
\usepackage{mathtools}
\usepackage{amsthm}
\usepackage[english]{varioref}
\usepackage{url}
\usepackage{multicol, multirow}
\usepackage{mparhack}
\usepackage{emptypage}
\usepackage{verbatim}
\usepackage{setspace}
\usepackage{dsfont}
\usepackage{rotating}
\usepackage[version=3]{mhchem}
\usepackage{bm}
\usepackage{enumitem}
\usepackage{tikz}
\usepackage{xcolor}
\usepackage{setspace}
\usepackage[abs]{overpic}
\usepackage{slashed}

\usepackage{hyperref}

\newcommand{\ahat}{\widehat{\alpha}}
\newcommand{\alphahat}{\widehat{\alpha}}
\newcommand{\ksihat}{\widehat{\xi}}
\newcommand{\xihat}{\widehat{\xi}}
\newcommand{\ttil}{\tilde{\tau}}
\newcommand{\ltil}{\tilde{l}}
\newcommand{\nastytwo}{C^{(2)}}

\newcommand{\gbil}{\mathcal{G}_{bil}}
\newcommand{\gquad}{\mathcal{G}_{quad}}
\newcommand{\up}[1]{^{(#1)}}
\newcommand{\upp}[2]{^{(#1),(#2)}}
\newcommand{\cC}{\mathcal{C}}

\renewcommand{\epsilon}{\varepsilon}
\renewcommand{\phi}{\varphi}

\newcommand{\zbar}{\overline{z}}

\newcommand{\rd}{{\mathrm d}}

\newcommand{\be}{\begin{equation}}
\newcommand{\ee}{\end{equation}}

\newcommand{\cO}{\mathcal{O}}
\newcommand{\cG}{\mathcal{G}}

\newcommand{\psibar}{\overline{\psi}}

\newcommand{\degav}[1]{\langle #1 \rangle}
\newcommand{\degsum}[1]{\llangle #1 \rrangle}

\theoremstyle{definition}

\theoremstyle{remark}

\theoremstyle{definition}

{\left\lbrace\begin{array}{@{}l@{}}}%
{\end{array}\right.}

\allowdisplaybreaks[2]

\usepackage[font=small,format=hang,labelfont={sf,bf}]{caption}

\makeatletter
\DeclareFontFamily{OMX}{MnSymbolE}{}
\DeclareSymbolFont{MnLargeSymbols}{OMX}{MnSymbolE}{m}{n}
\SetSymbolFont{MnLargeSymbols}{bold}{OMX}{MnSymbolE}{b}{n}
\DeclareFontShape{OMX}{MnSymbolE}{m}{n}{
    <-6>  MnSymbolE5
   <6-7>  MnSymbolE6
   <7-8>  MnSymbolE7
   <8-9>  MnSymbolE8
   <9-10> MnSymbolE9
  <10-12> MnSymbolE10
  <12->   MnSymbolE12
}{}
\DeclareFontShape{OMX}{MnSymbolE}{b}{n}{
    <-6>  MnSymbolE-Bold5
   <6-7>  MnSymbolE-Bold6
   <7-8>  MnSymbolE-Bold7
   <8-9>  MnSymbolE-Bold8
   <9-10> MnSymbolE-Bold9
  <10-12> MnSymbolE-Bold10
  <12->   MnSymbolE-Bold12
}{}

\let\llangle\@undefined
\let\rrangle\@undefined
\DeclareMathDelimiter{\llangle}{\mathopen}%
                     {MnLargeSymbols}{'164}{MnLargeSymbols}{'164}
\DeclareMathDelimiter{\rrangle}{\mathclose}%
                     {MnLargeSymbols}{'171}{MnLargeSymbols}{'171}
\makeatother
\usepackage[T1]{fontenc}
\usepackage{lmodern}

\newcommand{\NN}{\frac{1}{N}}

\title{The Analytic Bootstrap in Fermionic CFTs}
\author{Mark van Loon}
\affiliation{Mathematical Institute, University of Oxford, Andrew Wiles Building, Radcliffe Observatory Quarter, Woodstock Road, Oxford, OX2 6GG, UK}

\emailAdd{mark.vanloon@maths.ox.ac.uk}

\abstract{We apply the method of the large spin bootstrap to analyse fermionic conformal field theories with weakly broken higher spin symmetry. 
Through the study of correlators of composite operators, we find the anomalous dimensions and OPE coefficients in the Gross-Neveu model in $d=2+\epsilon$ dimensions and the Gross-Neveu-Yukawa model in $d=4-\epsilon$ dimensions, based only on crossing symmetry.
Furthermore a non-trivial solution in the $d=2+\epsilon$ expansion is found for a fermionic theory in which the fundamental field is not part of the spectrum.
The results are perturbative in $\epsilon$ and valid to all orders in the spin, reproducing known results for operator dimensions and providing some new results for operator dimensions and OPE coefficients.
}

\begin{document}
\noindent\today
\maketitle

\section{Introduction} \label{sec:introduction}
Conformal field theories, as fixed points of renormalization group flow, occupy a special place in the space of quantum field theories, and have been successfully used to describe a wide range of phenomena, such as boiling water at the critical point \cite{Schmidhuber:1997zz}.

The conformal bootstrap is a computational method whose main idea is to leverage the constraint of OPE associativity into non-perturbative statements.
In the most common approach one first expands a four-point function in terms of a basis of conformal blocks, which captures all the contributions from intermediate states in a particular channel, and one subsequently checks that crossing symmetry is satisfied, i.e. that the result is independent from the choice of expansion channel (for an alternative approach in Mellin space, see \cite{Gopakumar:2016cpb}).
Since the method only relies on crossing symmetry, it is genuinely non-perturbative and, since it is independent of a Lagrangian description, its results are very general.

An early success of the conformal bootstrap was to fully solve the minimal models in 2d CFT \cite{Belavin:1984vu}; however generalizing the method to higher dimensions proved very difficult, and the bootstrap lay dormant for many years.
In \cite{Rattazzi:2008pe} the bootstrap was revived and successfully applied to CFTs in dimension $d>2$, which kicked off the `numerical bootstrap', in which linear operators into $\mathbb{R}$ are applied to the crossing symmetry equation.
The existence or non-existence of linear operators with specific properties then constrains the spectrum.
We refer the interested reader to the excellent reviews \cite{Simmons-Duffin:2016gjk, Rychkov:2016iqz}.

There has also been increased interest in the `analytic bootstrap', in which the crossing symmetry equation is used to derive analytic results for CFTs.
For example, in \cite{Alday:2016njk} a method is given for studying CFTs at points of large twist degeneracy.
At this degenerate point, the contributions to a 4-point function $\cG^{(0)}(u,v)$ from operators around an accumulation point $\tau$ in the twist spectrum, are summed into `Twist Conformal Blocks' (TCBs) $H_\tau^{(0)} (u,v)$:
\begin{equation}
\cG^{(0)}(u,v) = \sum_\tau H_\tau ^{(0)}(u,v)\, .
\end{equation}
The theory is then perturbed by a small parameter $\epsilon$, which induces anomalous dimensions and OPE coefficient corrections, thereby breaking the twist degeneracy and changing the twist conformal blocks:
\begin{equation}
H^{(0)}_\tau(u,v) \rightarrow \sum_\rho B_{\tau,\rho} H^{(\rho)}_\tau(u,v) \, , \qquad \qquad \cG^{(0)}(u,v)  \rightarrow  \sum_{\tau,\rho} B_{\tau,\rho} H_\tau ^{(\rho)}(u,v)\, ,
\end{equation}
where $\rho$ measures powers of the spin in the breaking of the twist degeneracy.
These twist conformal blocks can be effectively calculated since they satisfy a recurrence relation
\begin{equation}
\mathcal{C} H^{(m+1)}_\tau(u,v) =  H^{(m)}_\tau(u,v)\, ,
\end{equation}
where $\mathcal{C} = \mathcal{C}_{\tau,d}$ is a differential Casimir operator.
Studying the analytic properties of the twist conformal blocks then constrains the spectrum of scaling dimensions and OPE coefficients of a wide variety of theories.

For example, an interesting result is found in \cite{Komargodski:2012ek, Fitzpatrick:2012yx}, where it is shown that if a CFT in $d>2$ has two operators with non-zero twists $\tau_1$ and $\tau_2$ respectively, then $\tau_1+\tau_2$ is an accumulation point of the twist spectrum of the CFT, i.e. there are infinitely many operators with twist arbitrarily close to $\tau_1+\tau_2$.\footnote{Recall that the twist $\tau$ of a primary operator of scaling dimension $\Delta$ and spin $l$ is defined as $\tau = \Delta - l$.}
This can be easily shown from the analytic properties of conformal blocks.
Consider two scalar bosonic operators $\phi_1, \phi_2$ of twists $\tau_1, \tau_2$, and study its four-point correlator $\langle \phi_1\phi_1\phi_2\phi_2 \rangle$, which satisfies the crossing relation
\begin{equation} \label{eq:introTwistAdd}
\cG_{2112}(u,v) = u^{\frac{\tau_1+\tau_2}{2}} v^{-\tau_1}\cG_{1122}(v,u) \, .
\end{equation}
On the right-hand side, the identity operator gives a contribution of $1$ to $\cG_{1122}(u,v)$, which implies the existence of a term \( u^{\frac{\tau_1+\tau_2}{2}}v^{-\tau_1}\) on the left-hand side.
Since conformal blocks \( G_{\tau,l}(u,v)\) behave like \( u^{\tau/2} \) for small \( u \), we see that there must be operators of approximate twist \( \frac{\tau_1+\tau_2}{2} \). 
Individual conformal blocks \( G_{\tau,l}(u,v)\) have only a logarithmic divergence in \( v \); therefore to produce the power-law divergence \( u^{\frac{\tau_1+\tau_2}{2}}v^{-\tau_1}\) on the left-hand side of equation \eqref{eq:introTwistAdd}, there in fact need to be an infinite number of operators whose twists accumulate at $\tau_1+\tau_2$.
This is just a taste of the powerful constraints that crossing symmetry imposes on CFTs.

Combined with any further constraints, such as exact conservation of the stress-energy tensor or conservation of currents associated to exact global symmetries, one may hope to fully constrain the spectrum order by order in $\epsilon$.
This was successfully done for several theories breaking higher spin symmetry in \cite{Alday:2016jfr}, for example reproducing results to first order in \( \epsilon \) in the $O(N)$ model at large \(N \), and in $\mathcal{N}=4$ super Yang-Mills.

All these computations rely on the specific form of conformal blocks, and as such mainly focus on scalar bosonic theories - or occasionally on correlators in supersymmetric theories where the superconformal primary is a boson.
Recently there have been some results in applying the bootstrap to 3d theories with fermions, in which universal numerical bounds on some operators were computed \cite{Iliesiu:2015qra, Iliesiu:2017nrv}.

In this paper we apply the analytic bootstrap to fermionic theories that are a perturbation of the theory of free Dirac fermions.
We study four-point functions of composite operators formed out of the fundamental fermions.
To the orders in the \(\epsilon\)-expansion to which we study these theories, their intermediate operators can be divided into bilinear operators (formed out of two fundamental fields) and quadrilinear operators (formed out of four fundamental fields).
Known anomalous dimensions of bilinear currents in the Gross-Neveu model in $d=2+\epsilon$ and in the Gross-Neveu-Yukawa model in $d=4-\epsilon$ are reproduced, and new results are found for OPE coefficient corrections of these bilinear currents, and for anomalous dimensions and OPE coefficient corrections of quadrilinear operators.
Furthermore a non-trivial solution in the \( d=2+\epsilon\) expansion is found for a fermionic theory in which anomalous dimensions scale logarithmically with the spin at first order in \( \epsilon \), which we conjecture to describe theories in which the fundamental field is not part of the spectrum.
The results are summarized in appendix \ref{app:results}.

The structure of this paper is as follows.
Section \ref{sec:setup} discusses the relevant background on crossing symmetry, twist conformal blocks and the free fermion theory.
Section \ref{sec:CrossAnalysis} introduces, in generality, the main method to study conformal field theories with weakly broken higher spin symmetry. 
These methods are then applied in sections \ref{sec:GN} and \ref{sec:GNY} to study the fermionic theories in the $d=2+\epsilon$ and $d=4-\epsilon$ expansions, paying special attention to the Gross-Neveu and Gross-Neveu-Yukawa models.

\section{Setup} \label{sec:setup}
In this section we recall some basic facts about CFTs, twist conformal blocks, and free fermions.

Recall that a CFT is completely determined by its `CFT data', i.e. the spectrum of primary operators $\cO_i$ of scaling dimension $\Delta_i$ and spin $l_i$, together with the OPE coefficients $c_{ijk}$, which specify the OPE algebra:
\begin{equation}
\cO_i \times \cO_j \sim \sum_k c_{ijk}\left(  \cO_k + \text{descendants} \right) \,.
\end{equation}

Consider the four-point function of four scalars $\phi_i$. 
Conformal invariance restricts the correlator to be of the following form \cite{Alday:2016mxe}: 
\begin{equation}
\langle \phi_i(x_1)\phi_j(x_2)\phi_k(x_3)\phi_l(x_4) \rangle = \left( \frac{x_{24}^2}{x_{14}^2} \right)^{\frac{\Delta_{ij}}{2}} \left( \frac{x_{14}^2}{x_{13}^2} \right)^{\frac{\Delta_{kl}}{2}} \frac{\cG_{ijkl}(u,v)}{x_{12}^{\Delta_i+\Delta_j}x_{34}^{\Delta_k+\Delta_l}}\, ,
\end{equation}
where $x_{ij} = x_i - x_j$, where $\Delta_{ij} = \Delta_i - \Delta_j$, and where $u,v$ are the standard conformal cross-ratios:
\begin{equation}
u= \frac{x_{12}^2 x_{34}^2}{x_{13}^2 x_{24}^2},\qquad v= \frac{x_{14}^2 x_{23}^2}{x_{13}^2 x_{24}^2}\, .
\end{equation}
This correlator may be evaluated by taking operator product expansions of $\cO_1 \times \cO_2$ and $\cO_3\times\cO_4$, or by expanding $\cO_1 \times \cO_4$ and $\cO_2 \times \cO_3$, with associativity of the OPE implying that both methods must give the same result.
This is called crossing symmetry, and yields the following relation:
\begin{equation} \label{eq:mixedcor}
v^{\frac{\Delta_j +\Delta_k}{2}} \cG_{ijkl}(u,v) = u^{\frac{\Delta_i + \Delta_j}{2}} \cG_{kjil}(v,u) \, .
\end{equation}
One can expand $\cG_{ijkl}(u,v)$ in terms of conformal blocks $G^{\Delta_{ij}, \Delta_{kl}}_{\tau,s} (u,v)$, which capture the contribution of a specific intermediate primary operator and all its descendants:
\begin{equation}
\cG_{ijkl} (u,v) = \sum_{\cO_{\tau,s}} c_{i j \cO_{\tau,s}} c_{k l \cO_{\tau,s}} G^{\Delta_{ij}, \Delta_{kl}}_{\tau,s} (u,v)\, ,
\end{equation}
where the sum is over primary operators $\cO_{\tau,s}$, of twist $\tau$ and spin $s$ that appear in the OPE of both $\phi_i \times \phi_j$ and $\phi_k \times \phi_l$, with OPE coefficients $c_{\bullet\bullet \cO}$.\footnote{Technically one needs operators $\cO_{\tau,s}$ and $\overline{\cO}_{\tau,s}$ that transform dually under any symmetry group, so that $\langle \cO_{\tau,s}\overline{\cO}_{\tau,s} \rangle \neq 0$.
That is, if they transform under representations $R$ and $\overline{R}$ of some symmetry group, one requires that the singlet representation satisfies $\mathbf{1} \subseteq R \otimes \overline{R}$. }

In the case of identical operators the expression reduces to
\begin{equation}
\cG(u,v) = \sum_{\tau,s} a_{\tau,s} G_{\tau,s} (u,v)\, ,
\end{equation}
where the sum is over operators $\cO_{\tau,s}$ of twist $\tau$ and spin $s$, $a_{\tau,s}$ is the squared OPE coefficient, and $G_{\tau,s}$ is the conformal block with identical external operators.
In a unitary theory the OPE coefficients are real and hence the $a_{\tau,s}$ are positive, a fact that is crucial to the numerical bootstrap program.

In the presence of a global symmetry group $G$ intermediate states will decompose into representations of $G$.
Suppose that the operators $\phi_i$ in the crossing relation \eqref{eq:mixedcor} transform in representations $R_i$ of the global symmetry group $G$.
Taking the OPE in the `direct' channel involves operators transforming in representations $R_D \subseteq R_i \otimes R_j$ and $\widetilde{R}_D \subseteq R_k \otimes R_l$, such that the singlet representation, under which the identity operator transforms, satisfies $\mathbf{1} \subseteq R_D \otimes \widetilde{R}_D $.
For intermediate operators in the `crossed' channel, one is interested in representations $R_C \subseteq R_k \otimes R_j$ and $\widetilde{R}_C \subseteq R_i \otimes R_l$ with $\mathbf{1} \subseteq R_C \otimes \widetilde{R}_C$.
In some common cases $R_D = \widetilde{R}_D$; however we shall encounter the case of the tensor product of two adjoint representations $\mathbf{(n^2-1)}$ of $U(n)$, which contains two unequal conjugate representations.
The relevant representations \( R_D\) and \( R_C \) may also be different in both channels, as is the case in the mixed correlators we shall consider later.

The conformal blocks are not known in general dimensions, but some exact results are known \cite{Dolan:2011dv}.
For example, the leading $u$-behaviour of the conformal blocks is known to all orders in $v$ in arbitrary dimensions.
Factoring the leading $u$-behaviour out of the conformal blocks:  $G^{\Delta_{12},\Delta_{34}}_{\tau,l}(u,v)  = u^{\tau/2} g^{\Delta_{12},\Delta_{34}}_{\tau,l}(u,v) $, they satisfy:
\begin{equation} \label{eq:leadingUBehaviour}
g^{\Delta_{12},\Delta_{34}}_{\tau,l}(u,v) =  \left(-\frac{1}{2}\right)^l (1-v)^l\,  _2 F_1\left(\frac{\tau}{2} + l -\frac{\Delta_{12}}{2},\frac{\tau}{2} + l +\frac{\Delta_{34}}{2},\tau+2l;1-v \right) +\cO(u) \,.
\end{equation}
Furthermore, the conformal blocks in 2d and 4d conformal blocks are known in closed form.
They are most easily expressed in the variables $z,\zbar$, which are related to $u,v$ through 
\begin{equation}
u = z \zbar\, , \qquad v = (1-z)(1-\zbar)\, .
\end{equation}
The 4d conformal blocks are then given by
\begin{equation}
G^{\Delta_{12},\Delta_{34}}_{\tau,l}(z,\zbar) = 
\left(-\frac{1}{2}\right)^l \frac{z \zbar}{z-\zbar} \left( k^{\Delta_{12},\Delta_{34}}_{\tau+2l}(z)k^{\Delta_{12},\Delta_{34}}_{\tau-2}(\zbar) - k^{\Delta_{12},\Delta_{34}}_{\tau-2}(z) k^{\Delta_{12},\Delta_{34}}_{\tau+2l}(\zbar) \right)\, ,
\end{equation}
and the 2d conformal blocks by\footnote{These are the conformal blocks associated to the global conformal group in $2d$ , i.e. they are not the Virasoro conformal blocks.}
\begin{equation}
G^{\Delta_{12},\Delta_{34}}_{\tau,l}(z,\zbar) = 
\left(-\frac{1}{2}\right)^l \left(1-\frac{1}{2} \delta_{0,l}\right)  \left( k^{\Delta_{12},\Delta_{34}}_{\tau+2l}(z)k^{\Delta_{12},\Delta_{34}}_{\tau}(\zbar) + k^{\Delta_{12},\Delta_{34}}_{\tau}(z) k^{\Delta_{12},\Delta_{34}}_{\tau+2l}(\zbar) \right)\, ,
\end{equation}
where
\begin{equation}
k^{\Delta_{12},\Delta_{34}}_{\beta}(x) = x^{\beta/2} \,\, _2 F_1\left(\frac{\beta}{2}-\frac{\Delta_{12}}{2},\frac{\beta}{2}+\frac{\Delta_{34}}{2},\beta,x \right)\, .
\end{equation}
From the definition of the $(z,\zbar)$ coordinates, it is clear that they provide a double covering of the $(u,v)$ coordinates, related by $z \leftrightarrow \zbar$.
Where appropriate, we make the choice of mapping the small $u$ limit onto the small $z$ limit, and the small $v$ limit onto the small $(1-\zbar)$ limit.

\subsection{Twist conformal blocks}
Consider a tree-level four-point function $\cG^{(0)}(u,v)$ of identical scalars $\phi$ at the point of large twist degeneracy, which can be decomposed into twist conformal blocks which capture the contributions of each degenerate twist in the spectrum:
\begin{equation} \label{eq:genTreeLevelDecomp}
\cG^{(0)}(u,v) = \sum_\tau H^{(0)}_\tau(u,v) = \sum_{\tau,l} a_{\tau,l} G_{\tau,l}(u,v) \, .
\end{equation}
Here we have assumed for simplicity that the external operators are identical; the definitions and properties in this section carry over in an obvious manner to the case of non-identical external operators.

The four-point function satisfies a crossing relation
\begin{equation}
v^{\Delta\up{0}_\phi} \cG\up{0}(u,v) = u^{\Delta\up{0}_\phi} \cG\up{0}(v,u)\, .
\end{equation}

We now turn on some small deformation away from the twist degenerate point, which we measure in a small parameter $\epsilon$, for example by turning on a coupling $g \sim \epsilon$.
We assume that the four-point function $\cG(u,v)$ admits the following expansion in terms of $\epsilon$:
\begin{equation}
\cG(u,v) = \cG\up{0}(u,v) + \epsilon \, \cG \up{1}(u,v) + \epsilon^2 \cG\up{2}(u,v) + \dots\, ,
\end{equation}
and that the twists and OPE coefficients  of intermediate operators can also be expanded in such powers:
\begin{align} \label{eq:corrExpansion}
\tau_l &= \tau_0 + \epsilon \,\gamma_{\tau_0,l}\up{1} + \epsilon^2\, \gamma_{\tau_0,l}\up{2} \dots \, , \\
a_{\tau,l} &= a_{\tau_0,l}\up{0} + \epsilon a_{\tau_0,l}\up{1}+ \epsilon^2 a_{\tau_0,l}\up{2} + \dots = a_{\tau_,l}\up{0} \left(1 + \epsilon \alpha_{\tau_0,l}\up{1} + \dots \right)\, .
\end{align}
We now use the result that the anomalous dimensions can be expanded in inverse powers of the conformal spin $J_{\tau,l}^2 = \left(l+\frac{\tau}{2}\right)\left(l+\frac{\tau}{2}-1\right)$:
\cite{Alday:2015eya, Caron-Huot:2017vep}
\begin{equation} \label{eq:gammaExp}
\gamma\up{m}_{\tau_0,l} = 2 \sum_\rho B\up{m}_{\tau_0,\rho} J_{\tau_0,l}^{-2\rho}\, ,
\end{equation}
where $\rho \in \mathbb{N}_0$, and where, by an abuse of notation, we refer to terms of the form $(\log^k J)/J^{2m}$ as $J^{-2\rho}$, where $(\rho)=  (m,\log^k J)$.
The same holds true for the \( \alphahat\up{m}_{\tau_0,l} \), which are shifted versions of the OPE coefficient corrections \( \alpha\up{m}_{\tau_0,l} \) (see section \ref{sec:CrossFurtherAnalysis} for the precise definition).
From the decomposition in equation \eqref{eq:genTreeLevelDecomp}, we can see how the various contributions to $\cG\up{1}(u,v)$ arise. 
For example, the corrections to the OPE coefficients create a correction to the correlator
\begin{equation}
\cG\up{1}(u,v) \supseteq \sum_{\tau_0,l} a_{\tau_0,l}\up{0} \gamma_{\tau_0,l}\up{1}  \frac{\partial}{\partial \tau}\bigg|_{\tau=\tau_0} G_{\tau,l}(u,v) \supseteq  \log u \sum_{\tau_0,\rho} B\up{1}_{\tau_0,\rho} H\up{\rho}_{\tau_0}(u,v)\, ,
\end{equation}
where we defined the twist conformal blocks
\begin{equation}
 H\up{\rho}_{\tau_0}(u,v) \equiv \sum_l a_{\tau_0,l}\up{0} J_{\tau_0,l}^{-2\rho}G_{\tau_0,l}(u,v) \, ,
\end{equation}
and where by `\( f \supseteq g \)' we mean that \( f \) contains terms of the form \( g \), i.e. \( f = g + \dots \) .
The conformal blocks satisfy an eigenvalue equation under the quadratic Casimir $\mathcal{D}_2$ of the conformal group \cite{Dolan:2011dv, Alday:2016njk,Hogervorst:2013kva} 
\begin{equation}
\mathcal{D}_2 G_{\tau,l}(u,v) = \left(l^2+l (\tau -1)+\frac{1}{2} \tau  (\tau -d)\right) G_{\tau,l}(u,v)\, .
\end{equation}
Introducing the shifted Casimir operator $\cC_{\tau,d} =\mathcal{D}_2  + \frac{1}{4}\tau(2d-\tau-2)$, the conformal blocks satisfy the eigenvalue equation $\cC_{\tau,d} G_{\tau,l}(u,v) = J^2_{\tau,l}  G_{\tau,l}(u,v)$, which in turn implies a recursion relation for the twist conformal blocks:
\begin{equation}
\cC_{\tau,d} H\up{m+1}_\tau(u,v) =  H\up{m}_\tau(u,v)\, .
\end{equation}
This is a differential equation that can be solved to find all the $H_\tau\up{m}$ iteratively once the tree-level result $H_\tau\up{0}$ is known.

The behaviour of the twist conformal blocks for small $u$ and $v$ is generally as follows:
\begin{equation} \label{eq:genTCBexp}
H\up{m}_\tau(u,v) \sim \frac{u^{\tau/2}}{v^{k_0}} v^m \left(h\up{m}_0(u) + h\up{m}_1(u) v + h\up{m}_2(u) v^2 + \dots \right)\, , 
\end{equation}
for some $k_0 \geq 0$, and where $h\up{m}_n(0) \neq 0$. 
This is consistent with the expectation that since $J^{-2}_{\tau,l} \sim l^{-2}$ for large $l$, the twist conformal blocks should become less divergent as one inserts more powers of $J^{-2}_{\tau,l}$.

As demonstrated in the introduction, the study of `enhanced divergences' in the crossing equation can prove very fruitful.
We shall define enhanced divergences to be terms $f(u,v)$ for which there exists an $n\in\mathbb{N}_{\geqslant 0}$ such that $\mathcal{C}_{\tau,d}^n \left( f(u,v)\right)$ has a power-law divergence in $v$, i.e. a divergence of the form $v^{-\beta}$, where $\beta > 0$. 
Specifically, this implies that they cannot be the sum of a finite number of conformal blocks.
Enhanced divergences therefore include all terms of the form $u^\bullet v^\beta$ where $\beta > 0$ is not integer, and $u^\bullet \log^k v$ where $k \geqslant 2$ is integer.

Generically all twist conformal blocks will possess enhanced divergences.
The type that they possess, depends on whether $k_0$, the tree-level $v$-divergence, is integer or not.
\begin{itemize}
\item If $k_0 \notin \mathbb{N}$ is not an integer, then all $H_\tau\up{m}(u,v)$ contain enhanced divergences from non-integer powers of $v$.
\item If $k_0 \in \mathbb{N}$ is integer, then $H_\tau\up{0}(u,v), \dots, H_\tau\up{k_0-1}(u,v)$ have enhanced divergences that are powers in $v$. 
The twist conformal blocks $H_\tau\up{k_0}(u,v),H_\tau\up{k_0+1}(u,v), \dots$ all develop enhanced divergences of the form $\log^2 v$.
This shall prove to be one of our most powerful tools, since such $\log^2 v$ terms can often only be produced on one side of the crossing equation.
\end{itemize}

\subsection{Free fermion CFT}
The theory of $N_f$ free massless Dirac fermions in $d$ dimensions has the action
\begin{equation}
S = \int \rd^d x~\psibar_i\slashed{\partial}\psi^i \,.
\end{equation}
From this it is clear that the fermions $\psi$ have scaling dimension $\Delta_\psi = \tfrac{d-1}{2}$, and that there is a global $U(N_f)$ symmetry, under which the fermions $\psi, \psibar$ transform in the fundamental, respectively anti-fundamental, representation.

The spectrum of bilinear primary operators in the theory consists of \cite{Giombi:2017rhm, Giombi:2016ejx}:
\begin{itemize}
\item A $U(N_f)$ scalar operator $\cO \equiv \psibar \psi = \psibar^i \psi_i$ of dimension $ \Delta_\cO = d-1$.
Spinor indices have been traced over, so that $\cO$ is a spacetime scalar.
\item A $U(N_f)$ adjoint operator $\cO^i_j = \psibar^i \psi_j - \frac{1}{N_f} \delta^i_j \psibar\psi $ of dimension $\Delta_A = d-1$, again with spinor indices traced over.
When suppressing $U(N_f)$ indices, we shall also refer to this operator as $\cO^A$.
\item A tower of totally symmetric conserved tensors $J_{\mu_1 \dots \mu_l} \sim \psibar \gamma_{\mu_l} \partial_{\mu_1}\cdots \partial_{\mu_{l-1}}\psi $  of dimension $\Delta_{J_l} = d-2+l$.
These correspond to the traceless symmetric representation of the $d$-dimensional Lorentz group, and the singlet representation of the global $U(N_f)$ symmetry.
Suppressing spacetime indices, we shall refer to the $U(N_f)$ singlet operators as $J_{S,l}$, and to the $U(N_f)$ adjoint operators as $J_{A,l}$.

The operator $J_{S,2}$ is the stress-energy tensor, while the operator $J_{A,1}$ is the global symmetry current.
\item A tower of mixed symmetry conserved tensors $B_{\mu_1 \cdots \mu_l\nu_1 \dots \nu_k} \sim \psibar \gamma_{\mu_1\nu_1\dots\nu_k} \partial_{\mu_2}\dots \partial_{\mu_l} \psi$ (with $1\leqslant k<d$), where $\gamma_{\mu_1\nu_1\dots\nu_k} = \gamma_{[\mu_1}\gamma_{\nu_1}\dots\gamma_{\nu_k]}$.
They correspond to representations of $SO(d)$ with highest weight $(l,1,\dots,1,0,\dots,0)$, i.e. to Young tableaux with $l$ boxes in the first row, and $k$ rows in total, and can transform in either the singlet or adjoint representation of $U(N_f)$.
They also saturate the unitarity bound $\Delta_{B_l} = d-2 +l$ \cite{Minwalla:1997ka}.
\end{itemize}

The two-point function of the fermions is as follows:
\begin{equation}
\langle \psibar^i (x_1) \psi_j (x_2) \rangle = \delta^{i}_j\frac{C_{\psi\psi}\slashed{x}_{12}}{(x^2_{12})^{\Delta_\psi+\tfrac12}}\,, \qquad  C_{\psi \psi} = \frac{\Gamma(\frac{d}{2})}{2\pi^{d/2} }\,,
\end{equation}
with the two-point functions $\langle \psi_i \psi_j \rangle$ and $\langle \psibar^i \psibar^j\rangle $ necessarily vanishing due to their spacetime representations.
With this normalization, all other correlators can be calculated from judicious use of Wick's theorem.

Our primary point of study will be the four-point correlators formed out of $\cO$ and $\cO^A$.
Using Wick contractions, and with the help of Mathematica (and the package `Gamma' \cite{Gran:2001yh}), the free theory result for $\langle \cO\cO\cO\cO \rangle$ can be found:
\begin{align}
&\langle \cO(x_1) \cO(x_2) \cO(x_3) \cO(x_4) \rangle = N^2C_{\psi\psi}^4 \frac{\cG(u,v)}{(x_{12}^2x_{34}^2)^{d-1}}\,,\\
&\cG(u,v) = 1 + \NN u^{\tfrac{d}{2}-1} \frac{(1-v)(1-v^{\tfrac{d}{2}})}{v^{\tfrac{d}{2}}} 
- \NN u^{\tfrac{d}{2}}\frac{1+v^{\tfrac{d}{2}}}{v^{\tfrac{d}{2}}} 
+ u^{d-1}\left(1+\frac{1}{v^{d-1}} - \NN \frac{1+v}{v^{\tfrac{d}{2}}} \right) + \NN u^d \frac{1}{v^{\tfrac{d}{2}}}\, ,
\label{eq:scalar4pt}
\end{align}
where $N \equiv N_f\,\text{Tr}(\mathbf{1})$, with the trace over spinor indices.
In the large $N$ limit, the disconnected diagrams should dominate.
Combining this with the fact that $\langle \cO(x_1) \cO(x_2) \rangle = N C_{\psi\psi}^2 (x_{12}^2)^{-(d-1)}$, the four-point function should in the large $N$ limit be that of a free boson of dimension $\Delta = d-1$, as is indeed the case.

In the four-point function of adjoints, there are multiple $U(N_f)$ tensor structures, arising from the possible representations of the exchanged intermediate operators.\footnote{
For $N_f \geqslant 4$ there are seven different irreducible representations: $(\mathbf{n^2-1}) \otimes (\mathbf{n^2-1}) = \bigoplus_{i=1}^7 R_i$.}
Therefore this correlator decomposes:\footnote{Two of the representations in the product of two $U(N_f)$ adjoints are conjugate representations and combine into one tensor structure.}
\begin{equation}
\langle \cO^{i_1}_{j_1}(x_1)\cO^{i_2}_{j_2}(x_2)\cO^{i_3}_{j_3}(x_3)\cO^{i_4}_{j_4} (x_4)\rangle \propto \sum_{k=1}^6 \cG_k(u,v) \left(\mathbf{T}_k\right)^{i_1\dots i_4}_{\,j_1\dots j_4}\,,
\end{equation}
where $\mathbf{T}_k$ is the tensor structure corresponding to the representation $R_k$ in the tensor product of two $U(N_f)$ adjoints.
We have also calculated all $\cG_k$, and shall give their properties in the main body when necessary.

Using the leading-$u$ behaviour of the conformal blocks from equation \eqref{eq:leadingUBehaviour}, we see that the exchanged operators at least contain operators of twist $\tau = d-2$, corresponding to the conserved currents, and operators of twist $\tau = 2d-2$, corresponding to quadrilinear operators $\cO_{quad} \sim \psibar\partial^{l_1}\psi\partial^{l_2}\psibar\partial^{l_3}\psi$ of spin \( l = l_1+l_2+l_3\).
The leading-$u$ behaviour can also be used to fix the squared OPE coefficients $a_{S,\tau,l}^{(0)}$ for these operators.
For example, the squared OPE coefficients $a_{S,d-2,l}^{(0)}$ of the bilinear currents $J_l$ in the OPE $\cO \times \cO$, are given by
\begin{equation} \label{eq:freeBilScalarOPE}
a_{S,d-2,l}^{(0)} = \left(1+(-1)^l\right)  \frac{2^l \Gamma \left(\frac{d}{2}+l-1\right)^2 \Gamma (d+l-2)}{N \Gamma \left(\frac{d}{2}\right)^2 \Gamma (l) \Gamma \left(d-3+2l\right)} \, ,  \qquad \qquad l \geqslant 2.
\end{equation}
Only operators of even spins appear, as necessary in the OPE of two identical operators. 

The question remains whether the $u^{d/2}$ and $u^d$ terms in equation \eqref{eq:scalar4pt} are the result of further operators appearing, or whether they arise from the sub-leading $u$-contributions in the conformal blocks $G_{d-2,l}(u,v)$ and $G_{2d-2,l}(u,v)$.
Checking in 2d and 4d, where we have closed form expressions for the conformal blocks, we find that the $u^{d/2}$ contribution is explained from the conformal blocks $G_{d-2,l}(u,v)$, while there needs to be an infinite tower of operators of twist $\tau = 2d-2+2n$ for $n\in \mathbb{Z}_{\geqslant 0}$ to account for all the terms of the form $u^{d-1+n}$.
This is a general feature of all the $\cG_k(u,v)$ encountered in this paper.

\section{Crossing analysis} \label{sec:CrossAnalysis}
In this section we perform a detailed analysis of the crossing equations, in its general form. 
We study a perturbation of the free fermion theory in $d_0>2$, in which no additional operators enter at first loop order.
Since the Gross-Neveu and Gross-Neveu-Yukawa theories we are interested in both violate one of these assumptions, this shall merely be a toy model to introduce the methods used in sections \ref{sec:GN} and \ref{sec:GNY}.
We restrict ourselves to studying the correlator $\langle\cO\cO\cO\cO\rangle$ because its analysis already contains the main ideas used in the paper.

Recall the free theory result in $d$ dimensions, equation \eqref{eq:scalar4pt}:
\begin{align}
\cG\upp{0}{d}(u,v) = 1 + \gbil\upp{0}{d}(u,v) + \gquad\upp{0}{d}(u,v) = 1 + H_{d-2}\upp{0}{d}(u,v) + \sum_{n=0}^\infty H\upp{0}{d}_{2d-2+2n}(u,v) \, , 
\end{align}
where the $\up{0}$ refers to the fact that this is the free theory, where $\up{d}$ indicates the dimension of space, and where
\begin{align}
\gbil\upp{0}{d}(u,v) &\equiv H_{d-2}\upp{0}{d}(u,v)
\qquad \,\,\,\,= \NN u^{\tfrac{d}{2}-1} \frac{(1-v)(1-v^{\tfrac{d}{2}})}{v^{\tfrac{d}{2}}} 
- \NN u^{\tfrac{d}{2}}\frac{1+v^{\tfrac{d}{2}}}{v^{\tfrac{d}{2}}}\, , \label{eq:freeBilTCB} \\
\gquad\upp{0}{d}(u,v) &\equiv  \sum_{n=0}^\infty H\upp{0}{d}_{2d-2+2n}(u,v) 
= u^{d-1}\left(1+\frac{1}{v^{d-1}} - \NN \frac{1+v}{v^{\tfrac{d}{2}}} \right) + \NN u^d \frac{1}{v^{\tfrac{d}{2}}}\, .
\label{eq:freeQuadTCB}
\end{align}

For clarity we shall first consider a situation in which the dimension $d$ of spacetime is unrelated to the small expansion parameter $\epsilon$, before adding the $\epsilon$-dependence that is encountered in the $d=2+\epsilon$ and $d=4-\epsilon$ expansions.

The external dimension of the operator $\cO$ is given by 
\begin{equation}
\Delta_\cO = d-1 + \gamma_\cO = d-1 + \epsilon\gamma\up{1}_\cO + \epsilon^2 \gamma\up{2}_\cO + \dots\, .
\end{equation}
Furthermore, we assume that the correlator can be expanded in terms of $\epsilon$:
\begin{equation}
\cG(u,v) = \cG\up{0}(u,v) + \epsilon \cG\up{1}(u,v) + \epsilon^2 \cG\up{2}(u,v) + \dots
\end{equation}
Now take the crossing equation, $v^{\Delta_\cO} \cG(u,v) = u^{\Delta_\cO} \cG(v,u)$, and expand it in powers of $\epsilon$:
\begin{align}
&v^{d-1}\left( 1+\epsilon \gamma\up{1}_\cO\log v + \epsilon^2\left(\gamma\up{2}_\cO \log v+\frac{1}{2}\left( \gamma\up{1}_\cO\right)^2 \log^2 v\right) +\dots\right) \nonumber \\
& \qquad\qquad\qquad\qquad \times\bigg(\cG\up{0}(u,v) + \epsilon \cG\up{1}(u,v) + \epsilon^2 \cG\up{2}(u,v)  + \dots \bigg)
=\, \Big(u \leftrightarrow v\Big)
\end{align}
Taking the results order by order in $\epsilon$, the crossing relation decomposes into a set of equations
\begin{align}
v^{d-1} \cG\up{0}(u,v) &= u^{d-1} \cG\up{0}(v,u) \label{eq:cross0thOrder} \\
v^{d-1}\left( \gamma\up{1}_\cO\log v \, \cG\up{0}(u,v) + \cG\up{1}(u,v)\right) &=
u^{d-1}\left( \gamma\up{1}_\cO\log u \, \cG\up{0}(v,u) + \cG\up{1}(v,u)\right)\label{eq:cross1stOrder}  \\
&\, \, \, \vdots \nonumber
\end{align}
Generally the crossing equations of lower order in \( \epsilon \) can be used to simplify the crossing equations of higher orders.
For example, substituting the order \(\epsilon^0\) equation \eqref{eq:cross0thOrder} into the order \( \epsilon^1\) equation \eqref{eq:cross1stOrder} simplifies the latter to
\begin{equation}\label{eq:cross1stOrder2} 
\gamma\up{1}_\cO\log v \, \cG\up{0}(u,v) + \cG\up{1}(u,v) =
\gamma\up{1}_\cO\log u \, \cG\up{0}(u,v) + \frac{u^{d-1}}{v^{d-1}} \cG\up{1}(v,u)\, .
\end{equation}

\subsection{Dimension shift}
In the theories in this paper the small parameter $\epsilon$ is related to the dimension in which the theory lives. 
Specifically the Gross-Neveu model lives in $d=2+\epsilon$, while the Gross-Neveu-Yukawa model lives in $d =4-\epsilon$ \cite{ZinnJustin:1991yn}.
In this case there will be corrections to the free theory correlators, OPE coefficients and scaling dimensions, entirely because of this dimensional shift.
In an interacting theory living in e.g. $d = d_0+\epsilon$, we want to define anomalous dimensions and OPE coefficient corrections with respect to the dimensions and OPE coefficients of the free theory in $d = d_0+\epsilon$ dimensions - and not with respect to those of the free theory in $d_0$ dimensions.
For computational purposes however, we will want to calculate twist conformal blocks in $d_0$ dimensions, so that we need to carefully keep track of the changes to twist conformal blocks from this dimensional shift.

For example, take the bilinear operators in the free fermion theory and consider a small change in dimension away from some fixed dimension: $d = d_0 \rightarrow d_0 + \epsilon$.
This changes the OPE coefficients and conformal blocks, leading to a change in the TCB:
\begin{align}
H_{d_0-2+\epsilon}^{(0),(d_0+\epsilon)}(u,v) = H_{d_0-2}^{(0),(d_0)}(u,v) + \epsilon \, \widetilde{H}_{d_0-2}^{(1),(d_0)} (u,v) + \mathcal{O}(\epsilon^2) \, .
\end{align}
In our analyses of the crossing equation, we shall always need to keep these terms in mind.
We shall now describe precisely the contributions to $\widetilde{H}_{d_0-2}^{(1),(d_0)} (u,v)$ resulting in this change.

The free theory OPE coefficients change because of their explicit dependence on the dimension $d$; therefore the change can simply be found by substituting $d = d_0+\epsilon$ and Taylor expanding.
The changes to the conformal blocks are twofold: firstly, the blocks are simply different in different dimensions (even if the twist and spin are the same), and secondly, the free theory twist $\tau = d-2$ of the intermediate operators depends on the dimension.

Let us capture these changes as follows:
\begin{align}
a_{d_0 + \epsilon-2,l}\up{0}
&= a_{d_0-2,l}\up{0}\left(1+\epsilon\, \widetilde{\alpha}_{d_0-2,l}\up{0} + \cO(\epsilon^2)\right) \, , \\
G_{d_0-2+\epsilon,l}^{(d_0+\epsilon)}(u,v) 
&=  G_{d_0-2,l}^{(d_0)}(u,v) + \epsilon \left. \frac{\partial}{\partial d} \right|_{d=d_0} G_{d_0-2,l}^{(d)}(u,v) + \epsilon \left. \frac{\partial}{\partial \tau} \right|_{\tau =d_0-2} G_{\tau,l}^{(d_0)}(u,v) + \cO(\epsilon^2) \, ,
\end{align}
so that
\begin{align} \label{eq:totalFreeCorrection}
\widetilde{H}_{d_0-2}^{(1),(d_0)} (u,v)
= \sum_{l=0}^\infty a_{d_0-2,l}\up{0}
 \left(\widetilde{\alpha}_{d_0-2,l}\up{0} G^{(d_0)}_{d_0-2,l}(u,v) + \left. \frac{\partial}{\partial \tau}\right|_{\tau =d_0-2} G^{(d_0)}_{\tau,l}(u,v) +  \left. \frac{\partial}{\partial d}\right|_{d=d_0} G^{(d_0)}_{d_0-2,l}(u,v) \right)\, .
\end{align}

In equation \eqref{eq:freeBilScalarOPE} the squared OPE coefficients $a_{d-2,l}\up{0}$ were given as a function of $d$.
By Taylor expanding this around $d= d_0$, the correction $\widetilde{\alpha}_{d_0-2,l}\up{0} $ can be found.

While we might also be able to compute $\left. \frac{\partial}{\partial \tau}\right|_{\tau =d_0-2} G^{(d_0)}_{\tau,l}(u,v)$ directly (for example in $d_0 = 2,4,6$, where the blocks are known in their full form), we are unable to compute $ \left. \frac{\partial}{\partial d}\right|_{d=d_0} G^{(d)}_{d_0-2,l}(u,v)$ since we do not know the (full) conformal blocks in arbitrary dimensions.
Instead we shall rely on an expansion of the free theory result in terms of the dimension to find the collective sum \eqref{eq:totalFreeCorrection}.

Since the correlator $\cG\upp{0}{d}(u,v)$ depends explicitly on the dimension $d$, we can simply set $d=d_0 + \epsilon$ and Taylor expand:
\begin{equation}
\cG\upp{0}{d_0+\epsilon}(u,v) = \cG\upp{0}{d_0}(u,v) + \epsilon\, \widetilde{\cG}\upp{1}{d_0}(u,v)+\cO(\epsilon^2)\, ,
\end{equation}
where
\begin{equation} \label{eq:dimCorrBil1}
\widetilde{\cG}\upp{1}{d_0}(u,v) = \log u \left( \frac{1}{2} \gbil\upp{0}{d_0} (u,v) +\cG\upp{0}{d_0}_{\text{quad}}(u,v) \right) - \log v \, \widehat{\cG}\up{d_0}(u,v)\, ,
\end{equation}
with $\widehat{\cG}\up{d}(u,v)$ satisfying the equation
\begin{equation}\label{eq:dimCorrBil2}
\frac{u^{d-1}}{v^{d-1}} \widehat{\cG}\up{d} (v,u) = 1 + \frac{1}{2} \gbil\upp{0}{d}(u,v)\, .
\end{equation}

Let us now finally turn to crossing.
The external operators $\cO$ have anomalous dimensions $\gamma_\cO$ defined with respect to the free theory in $d_0 + \epsilon$ dimensions:
\begin{equation}
\Delta_\cO= d-1+\gamma_\cO = d_0-1 +\epsilon + \gamma_\cO =   d_0-1 +\epsilon \left(\gamma_\cO\up{1} +1\right) + \cO(\epsilon^2)  \, .
\end{equation} 
The crossing equation takes the form
\begin{equation}
v^{d_0-1+\epsilon(\gamma_\cO\up{1} + 1) + \dots}\left( \cG\upp{0}{d_0}(u,v) + \epsilon\widetilde{\cG}\upp{1}{d_0}(u,v) + \epsilon\cG\up{1}(u,v) + \dots \right) = \big(u\leftrightarrow v\big)\, ,
\end{equation}
where $\cG\up{1}$ captures the corrections arising from the departure from the free theory.
Expanding this in $\epsilon$, and keeping only the first-order terms in $\epsilon$, yields the first order crossing equation
\begin{align} \label{eq:genCross1stOrder}
&\left( \gamma_\cO^{(1)}+1\right) \cG^{(0)}(u,v) \log v + \widetilde{\cG}^{(1)}(u,v)  + \cG^{(1)}(u,v)  \nonumber \\
&=\left( \gamma_\cO^{(1)}+1\right) \cG^{(0)}(u,v) \log u + \frac{u^{d_0-1}}{v^{d_0-1}}\widetilde{\cG}^{(1)}(v,u) + \frac{u^{d_0-1}}{v^{d_0-1}}\cG^{(1)}(v,u)\, ,
\end{align}
where all the $\cG$ are measured with respect to dimension $d_0$, and where the crossing equation for the free theory result, $v^{d_0-1} \cG\upp{0}{d_0}(u,v) = u^{d_0-1} \cG\upp{0}{d_0}(v,u)$, was used.

Plugging the results from equation \eqref{eq:dimCorrBil1} and \eqref{eq:dimCorrBil2} into this equation, we find that
\begin{align} \label{eq:bigCrossing1}
&\left( \gamma_\cO^{(1)}+1\right) \cG^{(0)}(u,v) \log v + \log u\left( \frac{1}{2} \gbil\up{0} (u,v) +\cG^{(0)}_{\text{quad}}(u,v) \right) - \log v \, \widehat{\cG} (u,v)  + \cG^{(1)}(u,v)  \nonumber \\
&=\left( \gamma_\cO^{(1)}+1\right) \cG^{(0)}(u,v) \log u + \frac{u^{d_0-1}}{v^{d_0-1}}\log v \left( \frac{1}{2} \gbil\up{0}(v,u) +\cG^{(0)}_{\text{quad}}(v,u) \right) &\nonumber\\
&\qquad\qquad\qquad\qquad\qquad\qquad\qquad\qquad\qquad- \log u  \left( 1+ \frac{1}{2} \gbil\up{0}(u,v) \right) +  \frac{u^{d_0-1}}{v^{d_0-1}}\cG^{(1)}(v,u) \, .
\end{align}
By direct computation one finds that the extra terms in equation \eqref{eq:bigCrossing1} (compared to equation \eqref{eq:cross1stOrder2}) all cancel.\footnote{An alternative quick way to see this, which only works at first order in $\epsilon$, is that the free theory, with $\gamma\up{1}_\cO = 0$ and $\cG\up{1} = 0$, should be a solution. 
After plugging this into equation \eqref{eq:bigCrossing1}, only the extra terms from the dimension shift remain, so that they must cancel against each other.}
Thus the first order crossing equation reduces to  \eqref{eq:cross1stOrder2}:
\begin{equation}\label{eq:cross1stOrder2Repeated} 
\boxed{ 
\gamma\up{1}_\cO\log v \, \cG\up{0}(u,v) + \cG\up{1}(u,v) =
\gamma\up{1}_\cO\log u \, \cG\up{0}(u,v) + \frac{u^{d_0-1}}{v^{d_0-1}} \cG\up{1}(v,u)\, .
}
\end{equation}
Note that we performed an expansion $d\rightarrow d_0+\epsilon$; in the case of an expansion $d\rightarrow d_0-\epsilon$, some of the signs in the intermediate equations would change, but equation \eqref{eq:cross1stOrder2Repeated} would remain the same. 

\subsection{Further analysis} \label{sec:CrossFurtherAnalysis}
To analyse the consequences of equation \eqref{eq:cross1stOrder2Repeated}, consider the effect of some intermediate operators of twist $\tau_0$ gaining a non-zero anomalous dimension $\gamma_{\tau_0,l}\up{1}$, or an OPE coefficient correction $\alpha_{\tau_0,l}\up{1}$ as per equation \eqref{eq:corrExpansion}.
This creates a correction to the first order correlator:
\begin{equation} \label{eq:genDfirstOrdCorr1}
\cG\up{1}(u,v)\supseteq\sum_{\tau_0,l} a_{\tau_0,l}\up{0}\left(\alpha_{\tau_0,l}\up{1} +\gamma_{\tau_0,l}\up{1} \partial_\tau\right) G_{\tau,l}(u,v)\, .
\end{equation}
From the form of the conformal blocks, we can deduce that this has a $\log u$ piece of the form 
\begin{equation}
\sum_{\tau_0,l} a_{\tau_0,l}\up{0} \gamma_{\tau_0,l}\up{1} \partial_\tau G_{\tau}(u,v) \Big|_{\log u} = \frac{1}{2} \sum_{\tau_0,l} a_{\tau_0,l}\up{0} \gamma_{\tau_0,l}\up{1} G_{d_0-2}(u,v) = \frac{1}{2}  \sum_{\tau_0, \rho}^\infty B_{\tau_0,\rho}\up{1} H_{\tau_0}\up{\rho}(u,v)\, .,
\end{equation}
where \( \gamma_{\tau_0,l}\up{1} \) was expanded in terms of the conformal spin as in equation \eqref{eq:gammaExp}.
The full $\log v$ part of equation \eqref{eq:genDfirstOrdCorr1} is hard to identify in general dimensions $d_0$.
Let us instead consider dimensions $d_0=2,4$ and focus only on enhanced divergent parts proportional to $\log v$, that is, terms of the form $\frac{\log v}{v^k}$ and $\log^m v$ for $k\geqslant 1, m \geqslant 2$.

We follow the arguments in \cite{Henriksson:2017eej}.
Firstly recall that the 2d and 4d conformal blocks take the special form:
\begin{align} 
\text{2d:}\,\, G^{\Delta_{12},\Delta_{34}}_{\tau,l}(z,\zbar) &= 
\left(-\frac{1}{2}\right)^l \left(1-\frac{1}{2} \delta_{0,l}\right)  \left( k^{\Delta_{12},\Delta_{34}}_{\tau+2l}(z)k^{\Delta_{12},\Delta_{34}}_{\tau}(\zbar) + k^{\Delta_{12},\Delta_{34}}_{\tau}(z) k^{\Delta_{12},\Delta_{34}}_{\tau+2l}(\zbar) \right) , \label{eq:gen2dBlocks} \\
\text{4d:}\,\,G^{\Delta_{12},\Delta_{34}}_{\tau,l}(z,\zbar)& = 
\left(-\frac{1}{2}\right)^l \frac{z \zbar}{z-\zbar} \left( k^{\Delta_{12},\Delta_{34}}_{\tau+2l}(z)k^{\Delta_{12},\Delta_{34}}_{\tau-2}(\zbar) - k^{\Delta_{12},\Delta_{34}}_{\tau-2}(z) k^{\Delta_{12},\Delta_{34}}_{\tau+2l}(\zbar) \right) .\label{eq:gen4dBlocks}
\end{align}
Let us now define 
\begin{equation}
\left(\ttil,\ltil\right) = \left(\tau, l+\frac{\tau}{2}\right)\, ,
\end{equation}
so that
\begin{equation}
\partial_\tau = \partial_{\ttil} + \frac{1}{2}\partial_{\ltil}\, , \qquad \partial_{l} = \partial_{\ltil} \, .
\end{equation}
It follows that $\partial_{\ttil} k^{\Delta_{12},\Delta_{34}}_{\tau+2l}(z) = 0$. We can therefore rewrite the $\partial_\tau$ part of equation \eqref{eq:genDfirstOrdCorr1} as follows:
\begin{align}  \label{eq:genDfirstOrdCorr2}
&\sum_{\tau_0,l} a_{\tau_0,l}\up{0}\gamma_{\tau_0,l}\up{1} \partial_\tau G_{\tau}(u,v) 
= \sum_{\tau_0,l} a_{\tau_0,l}\up{0}\gamma_{\tau_0,l}\up{1} \left(\partial_{\ttil} + \frac{1}{2}\partial_{\ltil}\right) G_{\tau,l}(u,v)  \nonumber \\
&= \sum_{\tau_0,l} a_{\tau_0,l}\up{0}\gamma_{\tau_0,l}\up{1} \partial_{\ttil}G_{\tau,l}(u,v)  - \frac{1}{2} \sum_{\tau_0,l} \partial_{\ltil} \left( a_{\tau_0,l}\up{0}\gamma_{\tau_0,l}\up{1}\right) G_{\tau,l}(u,v) + \sum_{\tau_0,l} \partial_{\ltil} \left( a_{\tau_0,l}\up{0} \gamma_{\tau_0,l}\up{1}G_{\tau,l}(u,v)  \right)\, .
\end{align}
The last term is a boundary term, so that it will not contain any enhanced divergences (see appendix \ref{app:boundaryTerm} for a more detailed discussion).
From the special form of the conformal blocks, \eqref{eq:gen2dBlocks} and \eqref{eq:gen4dBlocks}, the first term can be seen to not contain any enhanced divergences of the form $\frac{\log v}{v^k}$ or $\log^2 v$, since the tree-level twist conformal blocks do not contain such terms.

Plugging this back into equation \eqref{eq:genDfirstOrdCorr1}, we find that, ignoring the \( \log u \log v \) divergences:
\begin{equation} \label{eq:genDfirstOrdCorr3}
\sum_{\tau_0,l} a_{\tau_0,l}\up{0}\left( \alpha_{\tau_0,l}\up{1} +\gamma_{\tau_0,l}\up{1} \partial_\tau\right) G_{\tau,l}(u,v) \bigg|_{\log v,\, enh. div.} = \sum_{\tau_0,l} a_{\tau_0,l}\up{0} \ahat_{\tau_0,l}\up{1} G_{\tau,l}(u,v) \bigg|_{\log v,\, enh. div.}\, ,
\end{equation}
where $ \ahat_{\tau_0,l}\up{1}$ is defined by
\begin{equation} \label{eq:ahatdef}
 \widehat{\alpha}_{\tau_0,l}\up{1} = \alpha_{\tau_0,l}\up{1} - \frac{1}{2 a_{\tau_0,l}\up{0}}\partial_{l} \left(a_{\tau_0,l}\up{0}\gamma_{\tau_0,l}\up{1} \right) = \alpha_{\tau_0,l}\up{1} - \frac{1}{2} \partial_{l}  \gamma_{\tau_0,l}\up{1} - \frac{1}{2}\gamma_{\tau_0,l}\up{1} \partial_{l} \log a_{\tau_0,l}\up{0} \, .
\end{equation}
%

Let us now expand the anomalous dimensions and OPE coefficient corrections of the bilinears in terms of the conformal spin $J^2_{\tau,l}$, assuming there are no $\log J$ terms in the expansion of the anomalous dimensions:
\begin{equation}
\gamma_{\tau_0,l}\up{1} = 2 \sum_{m=0}^\infty  \frac{B_{\tau_0,m}\up{1}}{ J_{d-2,l}^{2m}}\, , \qquad\qquad \ahat_{\tau_0,l}\up{1} = 2 \sum_{m=0}^\infty A_{\tau_0,m}\up{1} \frac{\log J}{J_{d-2,l}^{2m}} + 2 \sum_{m=0}^\infty \frac{\widetilde{A}_{\tau_0,m}\up{1}}{ J_{d-2,l}^{2m}} \, .
\end{equation}

These will contribute to \( \log u \) terms in equation \eqref{eq:genDfirstOrdCorr1} as follows:
\begin{align} \label{eq:cross1stOrderLoguContr}
\left. \cG\up{1}(u,v) \right|_{v^{-\bullet}\log u} &\supseteq \sum_{m=0}^\infty B_{d_0-2,m}\up{1} H_{d_0-2}\up{m}(u,v)\, , \\
\left. \cG\up{1}(v,u) \right|_{u^{-\bullet}\log u} &\supseteq - \sum_{m=0}^\infty A_{d_0-2,m}\up{1} H_{d_0-2}\up{m}(v,u)\, .
\end{align}
Here we used the result $\left. H_\tau^{(k,\log J)}(u,v)\right|_{v^{-\bullet}\log v} = -\frac{1}{2} H_\tau\up{k}(u,v)\big|_{v^{-\bullet}}$, which follows from analytically continuing in $k$, identifying 
\begin{equation}
H_\tau^{(k_0,\log J)}(u,v) = -\frac{1}{2}\left. \frac{\partial}{\partial k}\right|_{k=k_0} H_\tau\up{k}(u,v)\, ,
\end{equation}
and the assumption that the $k$-dependence of $H_\tau\up{k}(u,v)$ takes the form in equation \eqref{eq:genTCBexp}.

To see how this can be used to constrain the anomalous dimensions and OPE coefficients, let us assume that the dimension satisfies $d_0>2$, so that there is a `gap' between $u^{\frac{d_0}{2}}$, the highest power of $u$ in the bilinear TCB, and $u^{d_0-1}$, the lowest power of $u$ in the quadrilinear TCBs.
This gap allows for the bilinear enhanced divergences, i.e. those of the form\footnote{If $d_0$ is an even integer, then not all such terms are enhanced divergences, since $v^{-\frac{d_0}{2}+m}$ may become regular in $v$.
Furthermore powers of $u$ can recombine with the quadrilinear TCBs. 
Special care needs to be taken in such dimensions, and in the definition of $h_{d_0-2}\up{m}(u,v)$ below, one needs to discard these terms.
}
\begin{equation}
\frac{u^{\frac{d_0}{2}-1}}{v^{\frac{d_0}{2}-m}}\,,\quad \frac{u^{\frac{d_0}{2}}}{v^{\frac{d_0}{2}-m}} \qquad \text{ for } m\in\mathbb{Z}_{\geqslant 0} \, ,
\end{equation}
to be studied without reference to the quadrilinear operators, since crossing maps the set of these bilinear divergences onto itself.
Define $h_{d_0-2}\up{m}(u,v)$ by
\begin{equation}
\left. H_{d_0-2}\up{m}(u,v)\right|_{enh. div.} = \frac{u^{\frac{d_0}{2}-1}}{v^{\frac{d_0}{2}}} h_{d_0-2}\up{m}(u,v)
\end{equation}
Focusing on the \( \log u \) part of equation \eqref{eq:genDfirstOrdCorr1}, and looking at these these divergences yields, after substitution of \eqref{eq:cross1stOrderLoguContr}:
\begin{equation}
\sum_m B_{d_0-2,m}\up{1} h_{d_0-2}^{(m)}(u,v)  = \gamma^{(1)}_\cO h^{(0)}_{d_0-2}(u,v)-\sum_m A_{d_0-2,m}\up{1} h_{d_0-2}^{(m)}(v,u)\, .
\end{equation}
From the tree-level result and our knowledge of the asymptotic behaviour of TCBs, we find that
\begin{equation}
\begin{cases}
h^{(0)}_{d_0-2}(u,v) &= 1-u-v \, , \\
h^{(m)}_{d_0-2}(u,v) &\sim u^0v^m \qquad \text{ for small } u, v \, .
\end{cases}
\end{equation}
Equating terms of orders $u^0v^0$, $u^mv^0$ and $u^0v^m$, we find that 
\begin{equation} \label{eq:genDimBilAnomAndOPE}
\boxed{ B_{d_0-2,0}\up{1} + A_{d_0-2,0}\up{1} = \gamma_\cO^{(1)}, \qquad B_{d_0-2,m}\up{1} = A_{d_0-2,m}\up{1} = 0\,  \text{ for } m \geqslant 1\, .}
\end{equation}
This result holds in perturbations of the free theory in which the intermediate operators do not change.
As we shall see in the Gross-Neveu-Yukawa model, there can be non-zero $B_{d_0-2,m}\up{1}$ if there is a coupling to new operators.

\subsection{Crossing in the presence of a global symmetry}
Consider the crossing equation for the correlator $\langle \phi_1\phi_2\phi_3\phi_4\rangle$ of four spacetime scalars that transform under representations $R_1,\dots R_4$ of some global symmetry group.

As mentioned in section \ref{sec:setup}, the intermediate operators in the `direct' channel transform under representations $R_D \subseteq R_1\otimes R_2$ and $\widetilde{R}_D\subseteq R_3\otimes R_4$ which satisfy $R_D \otimes \widetilde{R}_D \supseteq 1$.
Similarly in the `crossed channel' one encounters intermediate operators in representations $R_C \subseteq R_2\otimes R_3$ and $\widetilde{R}_C\subseteq R_1\otimes R_4$ which satisfy $R_C \otimes \widetilde{R}_C \supseteq 1$.

The crossing equations similarly decompose. 
The correlators in the two channels decompose as
\begin{equation}
\cG_{1234}(u,v) = \sum_i \cG_{R_{D,i}}(u,v) \mathbf{T}_{D,i}\, ,
\qquad
\cG_{3214}(v,u) = \sum_j \cG_{R_{C,j}}(v,u) \mathbf{T}_{C,j}\, ,
\end{equation}
where the $\mathbf{T}$ range over bases of tensor structures for the intermediate operators.
Crossing relates these two:
\begin{equation}
v^{\frac{1}{2}(\Delta_2+\Delta_3)} \sum_i \cG_{R_{D,i}}(u,v) \mathbf{T}_{D,i}
= u^{\frac{1}{2}(\Delta_1+\Delta_2)}\sum_j \cG_{R_{C,j}}(v,u) \mathbf{T}_{C,j}\, .
\end{equation}
By projecting either side onto the other's basis of tensor structures, one finds crossing relations of the form
\begin{equation}
v^{\frac{1}{2}(\Delta_2+\Delta_3)} \cG_{R_{D,i}}(u,v)
= u^{\frac{1}{2}(\Delta_1+\Delta_2)}\sum_j \beta_{ij} \cG_{R_{C,j}}(v,u)\, .
\end{equation}

Since we are interested in correlators of $U(N_f)$ singlets and adjoints, we consider the following tensor products of $U(n)$ representations:\footnote{We assume that $n\geqslant 4$. For $n<4$, the tensor product of two adjoint representations contains fewer representations. 
}
\begin{equation}
\mathbf{1}\otimes\mathbf{1} = \mathbf{1}\,, \qquad 
\mathbf{1} \otimes (\mathbf{n^2-1}) = (\mathbf{n^2-1})\,, \qquad 
(\mathbf{n^2-1})  \otimes (\mathbf{n^2-1})  = \bigoplus_{i=1}^7 R_i\, .
\end{equation}
Therefore the crossing relation for a mixed correlator such as $\langle \cO \cO \cO^A \cO^A \rangle$ always relates two different $\cG$ directly, while the crossing relation for the adjoint correlator $\langle \cO^A \cO^A \cO^A \cO^A \rangle$ is of the form
\begin{equation}
v^{\Delta_A} \cG_i(u,v) = u^{\Delta_A} \sum_j \beta_{ij}  \cG_j (v,u)\, .
\end{equation}
To describe the $R_i$, it is easiest to consider two operators $\cO^{i_1}_{j_1}$ and $\cO^{i_2}_{j_2}$ transforming in the adjoint representation of $U(n)$. 
Then the seven representations $R_i$ correspond to the following intermediate operators in the OPE $\cO^{i_1}_{j_1} \times \cO^{i_2}_{j_2}$ :
\begin{itemize}
\item The singlet representation, containing the singlet bilinear currents of even spin.
\item An adjoint representation containing operators that are symmetric under an interchange $(i_1 \leftrightarrow i_2)$ or $(j_1\leftrightarrow j_2)$.
This contains adjoint bilinear currents of even spin.
\item An adjoint representation containing operators that are antisymmetric under an interchange $(i_1 \leftrightarrow i_2)$ or $(j_1\leftrightarrow j_2)$.
This contains adjoint bilinear currents of odd spin.
\item Four representations containing the quadrilinears.
They consist of tensors with 4 indices, and can be classified according to their symmetry properties: $V^{(i_1 i_2)}_{(j_1 j_2)}, V^{(i_1 i_2)}_{[j_1 j_2]}, V^{[i_1 i_2]}_{(j_1 j_2)}, V^{[i_1 i_2]}_{[j_1 j_2]}$.
Note that the second and third representations in this list are conjugate representations.
\end{itemize}

\subsection{Finite-support solutions and analyticity in spin} \label{sec:CaronHuotFinSupp}
The study of enhanced divergences of twist conformal blocks as outlined above uses the assumption that the CFT data is analytic in the spin $l$.
However it is known that this analyticity can fail to hold for all spins: in this case we need to consider solutions with a finite support on the spin \cite{Alday:2016jfr, Heemskerk:2009pn}.

Recently it has been shown that the OPE coefficients and anomalous dimensions, under some mild assumptions regarding Regge behaviour in the theory, are in fact analytic in the spin all the way down to spin $l=2$ \cite{Caron-Huot:2017vep}, thereby limiting the finite support solutions to $l=0,1$. 
In the theories in this paper, operators of spin $l=0$ often do not appear in the correlators we consider, leaving only spin $l=1$ open to a finite support solution.

There is one caveat here: the argument in \cite{Caron-Huot:2017vep} shows that the CFT data for spin $l\geqslant 2$ is analytic in $l$ \emph{non-perturbatively}, while our analysis is perturbative in $\epsilon$.
Perturbatively one expects, from the violation of Regge behaviour, that the CFT data will be analytic in the spin down to some minimal spin $L$ proportional to the loop order.
We shall explicitly state in the rest of the paper whenever we use this result.

\subsection{Degeneracy} \label{sec:degeneracy}
It is possible that there are multiple operators with the same tree-level twist and spin, so that they enter the crossing equation on the same footing. 
That is, if there are different operators $\cO_i$ with twists $\tau_i = \tau_0 + \dots$ and OPE coefficients $a_{\tau,l,i}$, then they enter the crossing equation as
\begin{equation} \label{eq:degenCont}
 \sum_i a_{\tau,l,i} G_{\tau_i,l}(u,v)\, .
\end{equation}
In such a case our analysis does not find the CFT data of the individual operators, but a weighted average. 
For example, the sum in equation \eqref{eq:degenCont} has an $\epsilon \log u$ part equal to:
\begin{equation} \label{eq:degenContLogU}
\left. \sum_i a_{\tau,l,i} G_{\tau_i,l}(u,v)\right|_{\epsilon \log u} = \left(\sum_i a_{\tau_0,l,i}\up{0} \frac{\gamma_{\tau_0,l,i}\up{1}}{2} \right) G_{\tau_0,l}(u,v)  \, ,
\end{equation}
and as such, the crossing equation is only sensitive to the average\footnote{Theorems regarding analyticity or convergence of the OPE that rely on crossing symmetry, such as in \cite{Caron-Huot:2017vep}, generally apply to these averages.}
\begin{equation}
\degav{\gamma_{\tau_0,l}\up{1} } \equiv \frac{\sum_i a_{\tau_0,l,i}\up{0} \gamma_{\tau_0,l,i}\up{1}}{\sum_i a_{\tau_0,l,i}\up{0} } \,.
\end{equation}
We similarly define for any function the average $\langle f_{\tau_0,l} \rangle$ to be
\begin{equation} \label{eq:degavdef}
\degav{f_{\tau_0,l}} \equiv \frac{\sum_i a_{\tau_0,l,i}\up{0} f_{\tau_0,l,i}}{\sum_i a_{\tau_0,l,i}\up{0} } \,.
\end{equation}
For notational purposes, we shall also define the following sum over degenerate states:
\begin{equation} \label{eq:degsumdef}
\degsum{f_{\tau_0,l}} \equiv \sum_i f_{\tau_0,l,i}\,.
\end{equation}
Specifically, note that knowledge of $\degav{ f_{\tau_0,l} }$ does not determine $\degav{f_{\tau_0,l}^2 }$, a problem that we will need to consider in section \ref{sec:2dSecondOrder}.\footnote{For a different perspective: the $a_{\tau_0,l,i}\up{0}\geqslant 0$ can be considered as a probability distribution on the different operators of fixed twist $\tau_0$ and spin $l$, with $f_{\tau_0,l}$ a random variable. 
Knowledge of the first moment $\mathbb{E}\left[f_{\tau_0,l}\right]$ does not fix the second moment $\mathbb{E}\left[f_{\tau_0,l}^2\right]$; to fully determine the values $f_{\tau_0,l,i}$, or equivalently all moments $\mathbb{E}\left[f_{\tau_0,l}^n\right]$, one needs access to at least as many moments as there are degenerate operators.}

In our paper, this type of degeneracy is present for the quadrilinear operators in both models. 
Furthermore, in the Gross-Neveu-Yukawa model there is a degeneracy in the bilinear currents, which is resolved in section \ref{sec:GNYmixing} by considering multiple correlators simultaneously.

\section{The $d=2+\epsilon$ expansion and the Gross-Neveu model} \label{sec:GN}
In this section we study fermionic CFTs that weakly break higher spin symmetry, in dimension \( d= 2 + \epsilon \), order by order in \( \epsilon \). 
We pay particular attention to the (critical) Gross-Neveu model, which can be described by the following action
\begin{equation}
S = \int \rd^{2+\epsilon}x \left(\psibar^i \slashed{\partial}\psi_i + \frac{1}{2} g \psibar^i \psi_i \right)\, ,
\end{equation}
where $g \sim \epsilon$, so that at $\epsilon =0$ it reduces to the free fermion in $2$ dimensions.

Section \ref{sec:2dFirstOrder} discusses first order corrections to the CFT data of both the bilinear and quadrilinear operators in the singlet representation, and of bilinears in the adjoint representation.
Section \ref{sec:2dSecondOrder} discusses the second order corrections to the anomalous dimensions of bilinear currents in the Gross-Neveu model.

\subsection{First order}\label{sec:2dFirstOrder}
\subsubsection{Results}
To summarize the results in this section: through an analysis of the correlator $\langle \cO\cO\cO\cO\rangle$, we find that a highly non-trivial solution to the first-order crossing equation exists, which reduces to the Gross-Neveu model upon demanding that the first order anomalous dimensions do not scale logarithmically with the spin.
After demanding this, further results about the non-singlet operators are found through the study of the correlator \( \langle \cO^A\cO^A\cO^A\cO^A\rangle\) and mixed correlators such as \( \langle \cO^A\cO\cO\cO^A\rangle\).

The full solution for the singlets is of a similar form to that for 4d (bosonic) gauge theories studied in \cite{Henriksson:2017eej}.
We find that the singlet bilinear operators of even spin $l\geqslant 2$ have the following anomalous dimensions and OPE coefficient corrections:\footnote{\(S_r(n) \) denotes the $r$-th order harmonic number: \( S_r(n) = \sum_{k=1}^n \frac{1}{k^r} \).}
\begin{align}
\gamma_{S,0,l}\up{1} &= \beta\left( S_1(l-1) - 1 \right)\, ,\label{eq:2dBilAnom}\\
\ahat_{S,0,l}\up{1} &= \left( 2\gamma_\cO\up{1} +\beta\right) S_1(l-1) + \xihat_{-1}\, ,\label{eq:2dBilOPE}
\end{align}
where it is assumed that there is a unique twist 0, spin $2$ operator corresponding to the stress-energy tensor, and where $\xihat_{-1}$ is a constant related to the central charge.
The full OPE coefficient correction $\alpha_{S,0,l}\up{1}$ can then be found using the definition \eqref{eq:ahatdef}; we do not produce it here.

Almost all the $U(N_f)$ singlet quadrilinear operators are degenerate, and we find the following results for their infinite support solution:
\begin{align} 
\degav{\gamma_{S,\tau_0 ,l}^{(1),inf.}} &= \beta_{\tau_0} S_1\left(l+\frac{\tau_0}{2}-1\right) +\kappa_{\tau_0}\, ,\label{eq:2dQuadAnomForm} \\
\degav{\ahat_{S,\tau_0,l}^{(1),inf.}} &= \alphahat_{\tau_0} S_1\left(l+\frac{\tau_0}{2}-1\right)+ \ksihat_{\tau_0}\, ,\label{eq:2dQuadOPEForm}
\end{align}
where
\begin{align}
\beta_{\tau_0} &= - \frac{\eta}{N+\eta} \beta \, , \label{eq:2dQuadBeta}\\
\kappa_{\tau_0} &= 2 \gamma_\cO\up{1} \frac{N+2\eta}{N+\eta} + \frac{\eta \beta}{N+\eta} \left(2-S_1\left(\frac{\tau_0}{2} -1 \right) + \frac{1}{2} \delta_{\tau_0,2}\right) \, ,\label{eq:2dQuadKappa}\\
\ahat_{\tau_0} &= 2\gamma_\cO\up{1} + \frac{\eta \beta}{N+\eta} \left(1- 2 S_1\left(\frac{\tau_0}{2}-1\right) + S_1\left(\tau_0-2\right) + \frac{1}{4}\delta_{\tau_0,2} \right)  \, , \label{eq:2dQuadAlpha}\\
\ksihat_{\tau_0} &= \frac{\eta}{N+\eta} \left( \xihat_{-1} + \beta \ksihat_{\tau_0}\up{\beta} + \gamma_\cO\up{1} \ksihat_{\tau_0}\up{\gamma_\cO}  \right)  \, , \label{eq:2dQuadXi}
\end{align}
where we defined $\eta = (-1)^{\frac{\tau_0}{2}}$, and where, in equation \eqref{eq:2dQuadXi}:
\begin{align}
\ksihat_{\tau_0}\up{\beta} &= \zeta_2 + 3 S_1\left( \frac{\tau_0}{2}-1 \right)-S_1\left( \frac{\tau_0}{2}-1 \right)^2 - 2 S_1\left( \tau_0- 2 \right) \nonumber\\
& \qquad\qquad + S_1\left( \frac{\tau_0}{2}-1 \right)S_1\left( \tau_0- 2 \right) + \frac{1}{2} S_2 \left(\frac{\tau_0}{2}-1 \right) - \frac{5}{4}\delta_{\tau_0,2}\, , \label{eq:2dQuadXiBeta}  \\
\ksihat_{\tau_0}\up{\gamma_\cO} &= 6 S_1\left( \frac{\tau_0}{2}- 1 \right) - 4 S_1\left( \tau_0- 2 \right) - \delta_{\tau_0,2} + N\eta \left( 4 S_1\left( \frac{\tau_0}{2}- 1 \right) - 2 S_1\left( \tau_0- 2 \right) \right)\, ,\label{eq:2dQuadXiGamma}
\end{align}
where \(\zeta_2 = \zeta(2) = \frac{\pi^2}{6}\).

We find that the above solution with $\beta \neq 0$ requires the existence of a solution with finite support on the spin, i.e. with $\gamma\up{1}_l\neq 0$ only for \(l=0,\dots,L\). 
As per the results of \cite{Caron-Huot:2017vep}, we would expect \(L=1\) to first order in \(\epsilon\). 
Indeed such a solution exists, and for the $U(N_f)$ quadrilinear singlets it takes the form
\begin{align}
\degav{\gamma^{(1),fin.}_{S,\tau_0,0}} = \frac{N}{N+\eta} \frac{1}{\tau_0 -1} \left(\gamma_{fin} + \frac{\beta}{4N}\left(1-\delta_{\tau_0,2}\right) \right)\, ,
\end{align}
where \(\gamma_{fin} \) is a constant not fixed by our analysis.
The part of the solution that is independent of the infinite support solution, which is found by setting \( \beta =0 \), matches the form of the solutions found in \cite{Heemskerk:2009pn}.

\paragraph{Logarithmic scaling and results for Gross-Neveu\\}
Logarithmic scaling of the anomalous dimensions with the spin is known to occur in CFTs, for example at order \( \epsilon^3 \) in the Gross-Neveu model \cite{Giombi:2017rhm} or at order \( \epsilon^2 \) in the critical nonlinear sigma model in \( d=2+\epsilon \) \cite{Giombi:2016hkj}.
This behaviour can be understood from the fact that the nearly conserved currents of twist \( \tau = d-2 \) contribute a term of the form \( 1/{s^{d-2}}\) in the large spin expansion, which generates a logarithmic term in \(d = 2+\epsilon \).
However, we are unaware of any known theories in which such behaviour already occurs at first order in \( \epsilon \), and we would expect this to correspond to a theory in which \( \psi \) is not part of the spectrum, for example because it is prohibited by gauge symmetry.

Demanding that the first order anomalous dimensions do not scale logarithmically with the spin, which sets \( \beta=0\), reduces our results to those in the Gross-Neveu model in \( d = 2 + \epsilon \).
To motivate this, let us consider the analogous bosonic case in \( d=4\) dimensions.
A full analysis of those theories is performed in \cite{Henriksson:2017eej}, which studies the implications of crossing symmetry of the correlator \( \langle \phi^2\phi^2\phi^2\phi^2 \rangle \) for the CFT data.
Their results include the possibility of \( \log J \) terms in the anomalous dimensions, and in fact, they find a theory in which these appear, namely \( \mathcal{N} = 4 \) SYM.  
However in theories in which the fundamental field \( \phi \) appears, such as  the Wilson-Fisher model, an analysis like that of \cite{Alday:2016jfr} shows that no such terms may appear; the reason that the corrections in \( \mathcal{N}=4 \) SYM could have logarithmic behaviour is because gauge symmetry prevented the field \( \phi \) from appearing in the spectrum.
A similar thing is likely happening here, where a full analysis of the correlator \( \langle \psibar \psi \psibar \psi \rangle \) of fundamental fields may be able to conclude that \( \beta =0 \) if \( \psi \) is part of the spectrum.

When \( \beta =0 \) our results reduce to
\begin{align}
\gamma_{S,0,l}\up{1} &= 0\, ,\label{eq:2dGNBilAnom}\\
\alpha_{S,0,l}\up{1} &= 2\gamma_\cO\up{1}  S_1(l-1) + \xihat_{-1}\, .\label{eq:2dGNBilOPE}
\end{align}
Furthermore the quadrilinear operators have corrections of the form 
\begin{align} 
\degav{\gamma_{S,\tau_0 ,l}\up{1}} &= \kappa_{\tau_0}\, ,\label{eq:2dGNQuadAnomForm} \\
\degav{\ahat_{S,\tau_0,l}\up{1}} &= 2\gamma_\cO\up{1} S_1\left(l+\frac{\tau_0}{2}-1\right)+ \ksihat_{\tau_0}\, ,\label{eq:2dGNQuadOPEForm}
\end{align}
where
\begin{align}
\kappa_{\tau_0} &= 2 \gamma_\cO\up{1} \frac{N+2\eta}{N+\eta} \, ,\label{eq:2dGNQuadKappa}\\
\ksihat_{\tau_0} &= \frac{\eta}{N+\eta} \left( \xihat_{-1} + \gamma_\cO\up{1} \ksihat_{\tau_0}\up{\gamma_\cO}  \right)  \, , \label{eq:2dGNQuadXi}
\end{align}
with \( \ksihat_{\tau_0}\up{\gamma_\cO} \) as before.
To fix $\xihat_{-1}$, we use the relation for the OPE coefficient of an operator $\cO$ with the stress-energy tensor $T$ \cite{Komargodski:2012ek}:
\begin{equation} \label{eq:centralChargeRelation}
a_{\cO\cO T} = \frac{1}{c_T}\frac{d^2}{(d-1)^2} \Delta_\cO^2\, ,
\end{equation}
with $c_T$ the central charge of the theory.
Evaluating at $d=2+\epsilon$, with \( c_T\up{0} = N \), and subtracting the free theory correction to the OPE coefficient, one deduces that 
\begin{equation}
\alpha_{S,0,2}\up{0} = 2 \gamma_{\cO}\up{1}  - \frac{1}{N} c_T\up{1} = 2 \gamma_\cO\up{1}\, ,
\end{equation}
where the full central charge of the theory is  \( c_T =  c_T\up{0} + \epsilon  c_T\up{1} + \dots \), and where we used the result that the central charge corrections in the Gross-Neveu model only start at order \(\epsilon^3 \) \cite{Diab:2016spb}.

From these results for the singlet operators in the Gross-Neveu model, we deduce results for the non-singlet operators. 
Specifically, we find for the bilinear adjoint operators of even spin \( l \geqslant 2\):
\begin{align}
\gamma^{(1)}_{A,0,l} &= 0 \, , \\
\lambda_{AAl}^{(1)} &= \gamma_{\cO^A}^{(1)} S_1(l-1) + k_A \, , \\
\lambda_{SAl}^{(1)} &= \frac{1}{2}\left( \gamma_\cO^{(1)}+\gamma_{\cO^A}^{(1)} \right) S_1(l-1) + k_{SA} \, .
\end{align}
Here the \(\lambda_{\bullet\bullet \bullet} \up{1}\) are the (multiplicative) corrections to the (non-squared) OPE coefficients \(  c_{\bullet\bullet \bullet} \), i.e. they satisfy
\begin{equation}
c_{\bullet\bullet \bullet} = c_{\bullet\bullet \bullet}\up{0} + \epsilon c_{\bullet\bullet \bullet}\up{1} + \dots  = c_{\bullet\bullet \bullet}\up{0}\left( 1+ \epsilon \lambda_{\bullet\bullet \bullet} \up{1} + \dots \right)\, ,
\end{equation}
with \(c_{AAl} = c_{\cO^A\cO^A J_{S,l}} \) and \(c_{SAl} = c_{\cO\cO^A J_{A,l}} \).

Furthermore, we have found that for bilinear currents of odd spin $l$:
\begin{equation}
\gamma_{A,0,l}^{(1)} = 0 \, .
\end{equation}
The odd spin singlets cannot be directly accessed by our method since they do not appear as intermediate operators in the four-point correlators we consider.

The bilinear anomalous dimensions $\gamma^{(1)}_{S,0,l}$ and $\gamma^{(1)}_{A,0,l}$ match known results for the Gross-Neveu model in $2+\epsilon$ dimensions, found for example in \cite{Giombi:2017rhm}.

\subsubsection{Singlets - $\langle \cO\cO\cO\cO\rangle$} \label{sec:2dsinglets}
We first analyse the bilinear currents on their own and later add the quadrilinear operators to the analysis.

\paragraph{Bilinear currents\\}
From the general tree-level result, equation \eqref{eq:freeBilScalarOPE}, the tree-level squared OPE coefficients of the $U(N_f)$ singlet currents can be found in $d_0=2$:
\begin{equation} 
a_{S,0,l}^{(0)} = \left(1 + (-1)^l\right)  \frac{2^{ l}  \Gamma (l)^2}{N \Gamma \left(2l-1\right)}\, .
\end{equation}
We repeat here the first-order crossing equation \eqref{eq:cross1stOrder2Repeated}, and evaluate it at $d_0=2$: 
\begin{equation}\label{eq:cross1stOrder2D}
\gamma_\cO^{(1)}  \log v\, \cG^{(0)}(u,v) +  \cG^{(1)}(u,v) = \gamma_\cO^{(1)} \log u \,\cG^{(0)}(u,v) + \frac{u}{v}\, \cG^{(1)}(v,u) \,.
\end{equation}
As in section \ref{sec:CrossFurtherAnalysis}, corrections to the OPE coefficients and anomalous dimensions create a correction to the correlator:
\begin{equation} \label{eq:2dfirstOrdCorr}
\cG\up{1}(u,v)= \sum_{\tau_0,l} a_{\tau_0,l}\up{0}\left(\alpha_{\tau_0,l}\up{1} +\gamma_{\tau_0,l}\up{1} \partial_\tau\right) G_{\tau,l}(u,v)\, .
\end{equation}
When we expand the anomalous dimensions and OPE coefficient corrections in terms of the conformal spin $J_{0,l}^2 = l(l-1)$, these corrections will organize themselves in terms of the twist conformal blocks $H_0\up{\rho}(u,v)$; let us therefore investigate their analytical properties.
The free theory result is
\begin{equation}
H\up{0}_{0}(u,v) = \NN\left( \frac{(1-v)^2}{v}- u\frac{(1+v)}{v}\right)\, .
\end{equation}
This has enhanced divergences of the form $\frac{1}{v}$ and $\frac{u}{v}$, so that one expects $\log^2 v$ divergences in $H\up{1}_{0}(u,v), H\up{2}_{0}(u,v),\, \dots$. 
Indeed this is what happens: in terms of the $z,\zbar$ coordinates, for \( m \geqslant 1\), there is a term of the form 
\begin{equation}
H_0^{(m)}(z,\zbar) \supseteq  h^{(m)}(\zbar)\log(1-\zbar)^2\, , \qquad\qquad\quad h^{(m)}(\zbar) \sim (1-\zbar)^{m-1} \qquad \text{ for }(1-\zbar) \ll 1\,.
\end{equation}
To see the consequences of this, consider terms in the crossing equation \eqref{eq:cross1stOrder2D} of the form \(\log^2 v\).
These cannot be produced on the right-hand side, and hence must also be absent on the left-hand side. 
Consider now an expansion of \(\ahat_{0,l}\), which we recall encodes all the enhanced divergences proportional to \( \log v\) in equation \eqref{eq:2dfirstOrdCorr} (excluding \(\log u \log v \) divergences), in terms of the conformal spin $J^2=J^2_{0,l}$
\begin{equation}
\ahat_{0,l}\up{1} = \sum_{m=0}^\infty \left( A_{0,m}\up{1} \frac{\log J}{J^{2m}}+ \widetilde{A}_{0,m}\up{1} \frac{1}{J^{2m}} \right)\, .
\end{equation}
This contributes a term
\begin{equation}
\cG\up{1}(u,v)\supseteq \sum_{l} a_{0,l}\up{0}\alphahat_{0,l}\up{1} G_{0,l}(u,v) 
=  \sum_{m=0}^\infty \left( A_{0,m}\up{1} H_0\up{m,\log J}(u,v)+ \widetilde{A}_{0,m}\up{1} H_0\up{m}(u,v) \right) \, .
\end{equation}
We immediately conclude that \(A_{0,m}\up{1} = 0 \) for \( m \geqslant 1 \), since these would produce terms of the form \( \log^3 v \). 
Furthermore, note that a non-zero \(A_{0,0}\up{1}\) produces \( \log^2 v \) terms that can be cancelled by the \( \widetilde{A}_{0,m}\up{1} \) with \(m \geqslant 1 \). 
In fact, we find that this imposes that they are of the following form
\begin{equation}
\frac{\widetilde{A}_{0,1}\up{1}}{A_{0,0}\up{1}} = \frac{1}{6} \, ,\quad
\frac{\widetilde{A}_{0,2}\up{1}}{A_{0,0}\up{1}} = -\frac{1}{30} \, ,\quad
\frac{\widetilde{A}_{0,3}\up{1}}{A_{0,0}\up{1}} = \frac{4}{315} \, ,\quad
\frac{\widetilde{A}_{0,4}\up{1}}{A_{0,0}\up{1}} = -\frac{1}{105} \, , \quad \dots\,
\end{equation}
which the astute reader may recognize as the coefficients of the expansion of the harmonic number $S_1(l-1)$ in terms of the conformal spin $J^2=J^2_{0,l}$:
\begin{equation}
S_1(l-1) = \gamma_e + \log J + \frac{1}{6 J^2} - \frac{1}{30 J^4} + \frac{4}{315J^6} - \frac{1}{105J^8} + \dots\, ,
\end{equation}
with $\gamma_e$ the Euler-Mascheroni constant.
Hence we conclude that the general form of the OPE correction is 
\begin{equation}
\alphahat_{S,0,l}\up{1} = \alphahat_{-1} S_1(l-1) + \xihat_{-1}\, ,
\end{equation}
where \( \alphahat_{-1} \) and \( \xihat_{-1} \) are constants to be fixed.
Similarly we demand absence of terms of the form \(u^0 \log u \log^2 v\), which arise as follows:
\begin{equation} 
\cG\up{1}(u,v)\bigg|_{\log u}\supseteq
\sum_{l} a_{0,l}\up{0}\gamma_{0,l}\up{1} \partial_\tau\big|_{\tau=0} G_{\tau,l}(u,v)\bigg|_{\log u}
= \sum_{l} a_{0,l}\up{0}\frac{\gamma_{0,l}\up{1}}{2}G_{0,l}(u,v)\, .
\end{equation}
In the last line we used the fact that conformal blocks are of the form $G_{\tau,l}(u,v) = u^{\tau/2}g_{\tau,l}(u,v)$, with $g_{\tau,l}(u,v)$ analytic in $u$.
In the crossing equation, we then come to the same conclusion as for the \( \alphahat\up{1}_{0,l} \), i.e. that the anomalous dimensions are of the form:
\begin{equation}
\gamma_{S,0,l}\up{1} = \beta \left( S_1(l-1) -1 \right)\, ,
\end{equation}
where exact conservation of the stress-energy tensor was used to fix \(\gamma_{S,0,2}\up{1} = 0 \).

Let us now explicitly show how to fix the constant $\alphahat_{-1}$. 
We look at the $u^0 v^{-1}\log v$ part of the crossing equation:
\begin{equation} \label{eq:alphaFixer}
\gamma_{\cO}\up{1} \cG\up{0}(u,v) \bigg|_{\frac{u^0 \log v}{v}} + \cG\up{1}(u,v) \bigg|_{\frac{u^0\log v}{v}} = \frac{u}{v}\left( \cG\up{1}(v,u)\bigg|_{\frac{v^0 \log v}{u}} \right)\, ,
\end{equation}
and use the known expansions of $\alphahat_{S,0,l}\up{1}$ and $\gamma_{S,0,l}\up{1}$ to find that
\begin{align}
\cG\up{1}(u,v) \bigg|_{\frac{u^0 \log v}{v}} &= -\frac{1}{2} \alphahat_{-1} H_0\up{0}(u,v)\bigg|_{\frac{u^0}{v}} = - \frac{\alphahat_{-1}}{2N} \, ,\\ 
\cG\up{1}(v,u)\bigg|_{\frac{v^0 \log v}{u}} &= -\frac{1}{2} \beta H_0\up{0}(v,u)\bigg|_{\frac{v^0}{u}} = - \frac{\beta}{2N}\, .
\end{align}
Combining this with the tree-level result, we find that equation \eqref{eq:alphaFixer} reduces to \( \gamma_\cO\up{1} - \frac{1}{2} \alphahat_{-1} = -\frac{1}{2} \beta \), so that
\begin{equation}
\alphahat_{-1} = 2\gamma_\cO\up{1} + \beta\, .
\end{equation} 
At this point we can fix no further constants using our analysis.

\paragraph{Summation\\}
Given the form of the anomalous dimensions and OPE coefficient corrections, it is natural to consider the full bilinear part of the correction \eqref{eq:2dfirstOrdCorr}.
While the full correction can in fact be found, for our analysis we shall only need its leading-$u$ behaviour.
This is because on the left-hand side of the crossing equation, the infinite support CFT data of the quadrilinear operators is fully determined by the part that has an enhanced divergence of the form \( \frac{1}{v} \).
Since the crossing equation is of the form $\cG\up{1}(u,v) = \frac{u}{v} \cG\up{1}(v,u) + \dots$, on the right-hand side these terms are fully determined by the leading behaviour of corrections from the bilinear currents.

Following the notation in \cite{Henriksson:2017eej}, let us write the bilinear CFT data as
\begin{align}
\gamma_{S,0,l}\up{1} = \alpha_{1,1} S_1(l-1) + \alpha_{1,0} \, ,\\
\alphahat_{S,0,l}\up{1} = \alpha_{0,1} S_1(l-1) + \alpha_{0,0} \, .
\end{align}
The leading-$u$ behaviour of the correction to the four-point correlator will then be a function $f(u,v)$ that can be expanded as
\begin{equation}
f(u,v) = \alpha_{1,1} f_{1,1}(u,v) + \alpha_{1,0} f_{1,0}(u,v) + \alpha_{0,1} f_{0,1}(u,v) + \alpha_{0,0} f_{0,0}(u,v) \, .
\end{equation}
Explicitly summing, the functions are as follows:
\begin{align}
f_{1,1}(u,v) &= \left(\frac{-1+v^2}{4 N v}\right)\log u\log v  + \left(\frac{-1+v^2}{2 N v}\right)\mathrm{Li}_2(1-v) + \zeta_2 \frac{(1-v)^2}{2N v}\,, \\
f_{1,0}(u,v) &= \frac{(1-v)^2}{2N v}  \log u + \left( \frac{1-v}{2N} \right) \log v \, ,\\
f_{0,1}(u,v) &=  \frac{(1-v)^2}{2N v} \log v\, ,\\
f_{0,0}(u,v) &= \frac{(1-v)^2}{N v}\, ,
\end{align}
where \( \frac{(1-v)^2}{N v} \) is the leading-$u$ part of the tree-level $H_0\up{0}(u,v)$, and where \( \frac{-1+v^2}{2 N v} \) is the leading-$u$ part of \( H^{(0),harm}_0(u,v) \equiv \sum_l a_{S,0,l}\up{0} S_1(l-1) G_{0,l}(u,v)\).

\paragraph{Quadrilinears\\}
Recall that generally the quadrilinear operators are degenerate: there are multiple operators with the same twist and spin.
Let us define $\tau_0(n) = 2+2n$ for $n\geqslant 0$, and denote by $\langle \gamma_{n,l}\up{1} \rangle$ the (average) first-order anomalous dimension of the quadrilinear operators of twist $\tau_0(n)$ and spin $l$.\footnote{Since we only discuss quadrilinears that are  $U(N_f)$ singlets, we shall often omit the subscript `\(S\)', and for further ease of notation we write terms like $\gamma_{n,l}\up{1}$ to denote $\gamma_{S,\tau_0(n),l}\up{1}$.}
Then the free theory twist conformal blocks $H_{\tau_0(n)}\up{0}(u,v)$ can be found from the free theory correlator, equation \eqref{eq:freeQuadTCB}.
Decomposing these, the following formula can be found for the OPE coefficients:
\begin{equation} \label{eq:2dfreeQuadOPE} 
\degsum{a_{n,l}\up{0}} =\left(1+ \frac{(-1)^{n+1}}{N}\right)   \frac{ 2^{1+l}\Gamma (n+1)^2 \Gamma (l+n+1)^2}{\Gamma \left(2n+1\right) \Gamma \left(2l+2n+1\right)}\, ,
\end{equation}
where $\degsum{a_{n,l}\up{0}}$ is defined as in equation \eqref{eq:degsumdef}.
Like their bilinear counterpart, the quadrilinear TCBs have the property that for $m\geqslant 1$:
\begin{equation}
H_{\tau_0(n)}\up{m}(z,\zbar)\supseteq z^{\tau_0(n)/2} h\up{m}(\zbar) \log^2 (1-\zbar) \, , \qquad\qquad\quad h^{(m)}(\zbar) \sim (1-\zbar)^{m-1} \qquad \text{ for }(1-\zbar) \ll 1\,.
\end{equation} 
so that no linear combination of them is free of $\log^2 v$ divergences.
Another property they share with $H_0\up{0}(u,v)$ is that if one demands that the sum 
\begin{equation}
H\up{0,\log J}_{\tau_0(n)}(u,v) + \sum_{m=0}^\infty B\up{m}_{\tau_0(n)} H\up{m}_{\tau_0(n)}(u,v)
\end{equation}
is free of $\log^2 v$ divergences, then the $B\up{m}_{\tau_0(n)}$ must be coefficients in the expansion of the harmonic number $S_1(l+n)$ in terms of $J^{-2}_{\tau_0(n),l}$.
That is, the sum $\sum_l a_{n,l}\up{0} S_1(l+n) G_{\tau_0(n),l}(u,v)$ is free of $\log^2 v$ divergences.
From this we deduce that the CFT data of the quadrilinear operators must have the following form
\begin{align}
\degav{\gamma_{n,l}\up{1} }&= \beta_n S_1(l+n) +\kappa_n \, ,\\
\degav{\alphahat_{n,l}\up{1} }&= \alphahat_n S_1(l+n) + \xihat_n\, .
\end{align}
On the other hand, the quadrilinear TCBs have an interesting property not shared by their bilinear counterpart: namely, while their sum $\cG_{quad}\up{0}(u,v) = \sum_{n=0}^\infty H_{\tau_0(n)}\up{0}(u,v)$ is free of $\log v$ terms, each $H_{\tau_0(n)}\up{0}(u,v)$ contains $\log v$ terms, in such a way that only the following combination is free of $\log v$ terms:
\begin{equation}
\alpha_0 \sum_{n\in2 \mathbb{N}}H_{\tau_0(n)}\up{0}(u,v) + 
\alpha_1 \sum_{n\in2 \mathbb{N}+1}H_{\tau_0(n)}\up{0}(u,v)\, .
\end{equation}

This restricts the form that anomalous dimensions and OPE coefficients may take.
For example, since there can be no terms of the form \( \log u \log^2 v \) in the first-order crossing equation, we see that the anomalous dimensions must be of the following form:
\begin{equation} \label{eq:2dQuadAnomForm2}
\degav{ \gamma_{n,l}\up{1} } = 
\begin{cases}
\beta_0 S_1(l+n) +\kappa_n\,, & \qquad\text{ if } n \text{ even,} \\
\beta_1 S_1(l+n) +\kappa_n\,, & \qquad\text{ if } n \text{ odd.}
\end{cases}
\end{equation}

Recall the crossing equation, and consider the part that has a power-law divergence in $v$:
\begin{equation} \label{eq:2dFullDivergentCrossing}
\gamma_{\cO}\up{1} (\log v-\log u) \cG\up{0}(u,v) \bigg|_{\frac{1}{v}} + \cG\up{1}(u,v) \bigg|_{\frac{1}{v}} = \frac{u}{v}f(v,u)\, .
\end{equation}
By looking at specific terms in this equation order by order in \( u \), we are able to fix the CFT data of the quadrilinear operators.\footnote{For technical reasons, we used the coordinates \( (z,\zbar) \) in our computation, and looked at power-law divergent terms in \( (1-\zbar) \), order by order in \( z\).
This only affects some technical details of the computation: the idea is the same if one replaces \( \log u \mapsto \log z \) and  \( \log v \mapsto \log (1-\zbar) \) in the discussion below.
}
Specifically, we fix:
\begin{itemize}
\item \( \beta_n \) by looking at the \( \log u \log v \) part of equation \eqref{eq:2dFullDivergentCrossing}.
\item \( \kappa_n \) by looking at the \( \log u \) part of equation \eqref{eq:2dFullDivergentCrossing}. Note we do not need to consider the \( \log u\log v \) part, which has already been fixed by the \( \beta_n \).
\item \( \alphahat_n \) by looking at the \( \log v \) part of equation \eqref{eq:2dFullDivergentCrossing}. Again the \( \log u \log v \) part has already been fixed by the \( \beta_n \).
\item \( \xihat_n \) by looking at the remaining part of equation \eqref{eq:2dFullDivergentCrossing}, i.e. the terms without logarithms.
\end{itemize}
Doing this full computation, we find the results in equations \eqref{eq:2dQuadAnomForm}-\eqref{eq:2dQuadXiGamma}, which we do not reproduce here due to their length.
This completely fixes the infinite support solution in terms of three constants: \( (\beta,\gamma_\cO\up{1},c_T\up{1}) \), where we used the relation \eqref{eq:centralChargeRelation} to exchange dependence on \(\xihat_{-1} \) into dependence on the central charge correction \( c_T\up{1} \). 

Note that finite support solutions may exist.
Specifically, assuming the argument from section \ref{sec:CaronHuotFinSupp}, one would expect to find a finite support solution for operators of spin $l=0$.
We analyze this possibility by looking at the full \( \log u \log v \) part of the crossing equation, and truncating to a finite order by sending \( u \mapsto \delta u, v \mapsto \delta v\) and truncating in powers of \(\delta \).
Doing this, we find that such a finite support solution must exist, and it takes the form
\begin{align}
\degav{\gamma^{(1),fin.}_{S,\tau_0,0}} = \frac{N}{N+\eta} \frac{1}{\tau_0 -1} \left(\gamma_{fin} + \frac{\beta}{4N} \left( 1-\delta_{\tau_0,2}\right) \right)\, ,
\end{align}
for some undetermined constant $\gamma_{fin}$.

The part of this finite support solution that is independent of the infinite support solution, takes the same form as in \cite{Heemskerk:2009pn}, equation (4.13), after setting $\Delta$, the dimension of the external operator, to $1$, and writing \(\tau_0 = \tau_0(k)=2+2k\).
Similarly to the results in \cite{Heemskerk:2009pn}, our analysis shows that there are further finite support solutions in which the spin cutoff $L$ satisfies $L\geqslant 2$, which we would not expect to see given the analyticity results of \cite{Caron-Huot:2017vep}.

\subsubsection{Adjoints: $\langle \cO^{i_1}_{\,\, j_1}\cO^{i_2}_{\,\,  j_2}\cO^{i_3}_{\,\,  j_3}\cO^{i_4}_{\, \, j_4}\rangle$} \label{sec:2dadjoints}
Having determined the anomalous dimensions and OPE coefficients of the singlets, and having found the Gross-Neveu model as the theory in which \( \beta =0 \), let us focus on the anomalous dimensions of the adjoint currents by considering the four-point function of adjoint operators in the Gross-Neveu model.

Recall that the four-point function of adjoints decomposes according to the $U(N_f)$ representations of the intermediate operators, which are the representations occurring in the tensor product of two adjoints.
Let us denote the function corresponding to the symmetric adjoint representation by $\cG_{A,S}(u,v)$, and that of the antisymmetric adjoint by $\cG_{A,A}(u,v)$.
Let us also denote the function corresponding to the singlet representation by $\cG_S(u,v)$.

\paragraph{Even spin}
The bilinear part of $\cG^{(0)}_{A,S}(u,v)$, i.e. the conformal block $H_{d-2,AS} \up{0}(u,v)$, is, up to an overall normalization, the same as the conformal block $H_{d-2}\up{0}(u,v)$ in the correlator $\langle \cO \cO \cO \cO \rangle$.

The crossing relation is of the form
\begin{equation}
v^{\Delta_A} \cG_{A,S} (u,v) = u^{\Delta_A}  \sum_{j=1}^6 \beta_{2j} \cG_j (v,u)\, ,
\end{equation}
where the $\beta_{ij}$ are constants fixed by the representation theory of $U(N_f)$.

Analysing crossing to first order in $\epsilon$ yields
\begin{equation}
\gamma_A^{(1)} \log v \, \cG_{A,S}^{(0)}(u,v) + \cG_{A,S}^{(1)}(u,v) = \gamma_A\up{1} \log u \, \cG_{A,S}^{(0)}(u,v) + \frac{u}{v} \sum_{j=1}^6 \beta_{2j} \cG_j ^{(1)}(v,u)\, ,
\end{equation}
where crossing was used in its tree-level form.
Since the tree-level result is the same as for the bilinear twist conformal block in section \ref{sec:2dsinglets}, we conclude that the expansion of $\gamma_{A,l}\up{1}$ must be constant to ensure that there is no $\log u\log^2 v$ term in the crossing equation:
\begin{equation}
\boxed{
\gamma_{A,0,l}^{(1)} = \gamma_{A,0,2}^{(1)} }\qquad \qquad l \geqslant 2 \quad \text{even}\,. 
\end{equation}

\paragraph{Odd spin}
The bilinear part of $\cG^{(0)}_{A,A}(u,v)$ is different from that in the correlator $\langle \cO\cO\cO\cO \rangle$: up to an overall normalization, we find it to be, in general dimension $d$:
\begin{equation}\label{eq:oddSpinTCB}
H_{d-2,AA} \up{0}(u,v) = u^{\frac{d}{2}-1}
\frac{(1-v)\left(1+v^{\frac{d}{2}}\right)}{v^{\frac{d}{2}}}
 + u^{\frac{d}{2}} \frac{1-v^{\frac{d}{2}}}{v^{\frac{d}{2}}}\, ,
\end{equation}
so that in $2$ dimensions
\begin{equation}\label{eq:oddSpinTCB2d}
H_{0,AA} \up{0}(u,v) = 
\frac{1-v^2}{v}  + u\frac{1-v}{v}\, .
\end{equation}
The crossing relation is of the form
\begin{equation}
v^{\Delta_A} \cG_{A,A} (u,v) = u^{\Delta_A}  \sum_{j=1}^6 \beta_{3j} \cG_j (v,u)\, .
\end{equation}
The crossing relation at first order in $\epsilon$ gives a very similar result to that for $\cG_{A,S}$:
\begin{equation}
\gamma_A^{(1)} \log v\, \cG_{A,A}^{(0)}(u,v) + \cG_{A,A}^{(1)}(u,v) = \gamma_A \log u \, \cG_{A,A}^{(0)} + \frac{u}{v}\sum_{j=1}^6 \beta_{3j} \cG_j ^{(1)}(v,u)\,.
\end{equation}
From the explicit form of the twist conformal block $H\up{0}_{0,AA}(u,v)$ in equation \eqref{eq:oddSpinTCB2d}, all higher TCBs $H^{(m)}_{0,AA}(u,v)$ can be calculated.
We find that for $m \geqslant 1$, they have \( \log^2(1-\zbar) \) contributions that are precisely the same as for the even spin case; the standard argument of demanding no $\log u \log^2 v$ divergences then implies that the $\gamma_{A,0,l}^{(1)}$ are constant for odd $l$.

Whether this constant can be non-zero depends on the existence of solutions with finite support on the spin.\footnote{Note that an expansion in inverse powers of the conformal spin $J_{0,l}^2 = l(l-1)$ makes no sense at $l=1$.}
Since the solution in the quadrilinears can be unbounded (in both spin and twist), we cannot rule out this possibility easily, and we shall revisit this issue in section \ref{sec:GNfinsupp}.
For now, we can only make the following ansatz:
\begin{equation}
\gamma_{A,0,l}^{(1)} = \omega_\infty +\sum_{m=1}^L \omega_m \delta_{k,l} \, , \qquad \qquad l\geqslant 1 \text{ odd} .
\end{equation}
Imposing exact conservation of the global symmetry current, which is a spin 1 current in the $U(N_f)$ adjoint representation, then shows that $\omega_\infty +\omega_1 = 0$.
We shall show in section \ref{sec:GNfinsupp} that there is no such finite support solution, so that in fact
\begin{equation}
\boxed{
\gamma_{A,0,l}\up{1} = 0 \, , \qquad l\geqslant 1 \text{ odd} .
}
\end{equation}

\subsubsection{Mixed correlators} \label{sec:2dMixedCorrs}
There is a plethora of mixed four-point correlators to consider:
\begin{align}
\langle \cO^A \cO \cO \cO^A \rangle \qquad&\rightarrow \qquad\cG_A(u,v) \\
\langle \cO \cO^A \cO^A \cO \rangle \qquad&\rightarrow \qquad\cG'_A(u,v) \\
\langle \cO \cO \cO^A \cO^A \rangle \qquad&\rightarrow \qquad\cG_S(u,v) \\
\langle \cO \cO^A \cO \cO^A \rangle \qquad&\rightarrow \qquad\widetilde{\cG}_A (u,v)
\end{align}

In the free theory all these $\cG$ have the same bilinear contribution. 
Using the property of conformal blocks \cite{Dolan:2011dv}:
\begin{equation}
G^{\Delta_{ij},\Delta_{kl}}_{\tau,s} (u,v) = v^{\frac{\Delta_{ij}-\Delta{kl}}{2}}  G^{-\Delta_{ij},-\Delta_{kl}}_{\tau,s} (u,v) \,, \label{eq:mixedconfblockid}
\end{equation}
and the fact that only even spin bilinear operators appear in the relevant OPEs, it can be shown that to all orders in $\epsilon$:
\begin{align}
 v^{\Delta_\cO} \left. \cG_A(u,v)\right|_{bil.} &= \left. v^{\Delta_A} \cG'_A (u,v)\right|_{bil.} \label{eq:mixedconsist} \,.
\end{align}

Since the external dimensions are no longer identical, there will be extra contributions to the conformal blocks:
\begin{equation}
G^{\Delta_{AS}, \Delta_{SA}}_{\tau,l} (u,v) = G^{0,0}_{\tau_0,l} (u,v) + \epsilon \, \widetilde{G}_{\tau_0,l}(u,v) + \epsilon\, G'_{\tau_0,l}(u,v)+ \dots \,, 
\end{equation}
where $\widetilde{G}_{\tau_0,l}$ is the correction due to the fact that the dimensions $\Delta_{AS}\equiv \Delta_A-\Delta_\cO$ and $ \Delta_{SA} = -\Delta_{AS}$ in $G^{\Delta_{AS}, \Delta_{SA}}_{\tau,l}$ may get corrections at order $\epsilon$, and where \( G'_{\tau_0,l}(u,v)  \) captures the changes to the blocks due to a dimensional correction and possibly non-zero anomalous dimensions.
Note that the following holds for the correction of  $G^{\Delta_{SA}, \Delta_{AS}}_{\tau,l} = G^{-\Delta_{AS}, -\Delta_{SA}}_{\tau,l}$:
\begin{equation}
G^{\Delta_{SA}, \Delta_{AS}}_{\tau,l} (u,v) = G^{0,0}_{\tau_0,l} (u,v) - \epsilon\,  \widetilde{G}_{\tau_0, l} (u,v) + \epsilon\, G'_{\tau_0,l}(u,v)+ \dots \, .
\end{equation}
Expanding equation \eqref{eq:mixedconsist} to order $\epsilon$, we find that:
\begin{align}
& v\left(1+\epsilon \gamma_\cO^{(1)} \log v + \dots\right) \sum_{l}  a_{SAl}^{(0)} \left( 1+ \epsilon \alpha_{SAl}^{(1)} + \dots \right) \left( G_{0,l}^{0,0}(u,v) + \epsilon \,\widetilde{G}_{0,l}(u,v) + \epsilon \, G'_{0,l}(u,v) +\dots \right) \nonumber \\
=\,\, & v\left(1+\epsilon \gamma_A^{(1)} \log v + \dots\right) \sum_{l}  a_{SAl}^{(0)} \left( 1+ \epsilon \alpha_{SAl}^{(1)}+ \dots \right) \left( G_{0,l}^{0,0} (u,v)- \epsilon \,\widetilde{G}_{0,l}(u,v) + \epsilon \, G'_{0,l}(u,v) +\dots \right)\, .
\end{align}
Matching the order $\epsilon$ part, we find that
\begin{align}
\sum_{l}  a_{SA,l}^{(0)}\widetilde{G}_{0,l}(u,v) &= \frac{\gamma_A^{(1)} - \gamma_{\cO}^{(1)}}{2} \log v \sum_{\tau_0,l}   a_{SAl}^{(0)} G^{0,0}_{0,l}(u,v) = \frac{\gamma_A^{(1)} - \gamma_{\cO}^{(1)}}{2} \log v \, H\up{0}_{0,A}(u,v) \, , \label{eq:1stordermixedconsist} 
\end{align}
an identity that shall prove useful later.

\paragraph{Crossing for $\langle \cO \cO^A \cO \cO^A \rangle$\\}
The crossing relation for $\widetilde{\cG}_A(u,v)$ maps it onto itself:
\begin{equation} \label{eq:tGAtGACrossing}
v^{\frac{\Delta_\cO+\Delta_A}{2}} \widetilde{\cG}_A(u,v) = u^{\frac{\Delta_\cO+\Delta_A}{2}} \widetilde{\cG}_A(v,u) 
\end{equation}
This makes the analysis slightly simpler; however, just like in the crossing relation for $\langle \cO^A \cO \cO \cO^A \rangle$, the conformal blocks have unequal external dimensions.
The relevant conformal blocks are 
\begin{equation}
G^{\Delta_{SA},\Delta_{SA}}_{\tau,l} (u,v) = G^{0,0}_{\tau,l} (u,v) + \epsilon \, \widetilde{F}_{0,l} (u,v) +\dots\, ,
\end{equation}
where $\widetilde{F}$ is the correction due to the fact that $\Delta_{SA}$ may acquire an anomalous dimension at order $\epsilon$.
However the identity \eqref{eq:mixedconfblockid} applied to $G^{\Delta_{SA},\Delta_{SA}}_{\tau,l} (u,v) = G^{\epsilon \gamma_{SA}, \epsilon \gamma_{SA}}_{\tau,l} (u,v) + \dots$ implies that 
\begin{equation}
G^{\epsilon \gamma_{SA}, \epsilon \gamma_{SA}}_{\tau,l} (u,v) 
=G^{-\epsilon \gamma_{SA},- \epsilon \gamma_{SA}}_{\tau,l} (u,v)\, ,
\end{equation}
so that the corrections due to the external dimensions must be an even function in $\epsilon$, thus forcing $\widetilde{F}$ to vanish.
We may therefore ignore the added subtleties of different external dimensions, and simply get the first order crossing equation
\begin{equation}\label{eq:mixCross1stOrder2D}
\frac{\gamma_\cO^{(1)} + \gamma_{A}\up{1} }{2}  \log v\, \widetilde{\cG}_A^{(0)}(u,v) +  \widetilde{\cG}_A^{(1)}(u,v) 
= \frac{\gamma_\cO^{(1)} + \gamma_{A}\up{1} }{2} \log u \,\widetilde{\cG}_A^{(0)}(u,v) + \frac{u}{v}\, \widetilde{\cG}_A^{(1)}(v,u) \,.
\end{equation}
From the standard argument of being free of $\log^2 v$, we find that the bilinear anomalous dimensions are all constant.\footnote{Harmonic number terms can be excluded by the crossing equation for \( \langle \cO^A\cO\cO\cO^A \rangle\), since they would imply the presence of harmonic number terms in the singlet operators.} 
Projecting onto the $u^0  v^{-1} \log v$ part of the equation, so as to isolate the bilinear operators, we find that 
\begin{align} 
\frac{\gamma_A^{(1)} + \gamma_\cO^{(1)} }{2} \, \left. \widetilde{\cG}^{(0)}_A(u,v)\right|_{u^0  v^{-1}} + \left. \sum_l a_{SAl}\up{0}\ahat_{SAl}\up{1} G_{0,l}^{0,0}(u,v)  \right|_{u^0  v^{-1}\log v} 
&=  \frac{u}{v} \frac{\gamma_{A,0,2}^{(1)}}{2} \left. \widetilde{H}_{0,A}\up{0}(v,u)\right|_{v^0 u^{-1} } \nonumber \\
&= \frac{u}{v} \frac{\gamma_{A,0,2}^{(1)}}{2} \left. \widetilde{\cG}^{(0)}_A(u,v)(v,u)\right|_{v^0 u^{-1} } \, . \label{eq:tGAtGAcrossres}
\end{align}

\paragraph{Crossing for $\langle \cO^A \cO \cO \cO^A \rangle$\\}
Consider the crossing equation relating $\cG_A$ and $\cG_S$:
\begin{equation}
v^{\Delta_\cO} \cG_A(u,v) = u^{\frac{\Delta_\cO+ \Delta_A}{2}} \cG_S(v,u)\, ,
\end{equation}
and expand it to order $\epsilon$ to find
\begin{equation} \label{eq:mixedcrossA1}
v\left( \gamma_\cO^{(1)} \log v \, \cG^{(0)}_A(u,v) + \cG^{(1)}_A(u,v) \right)  
=u \left( \frac{\gamma_\cO^{(1)}+\gamma^{(1)}_A}{2} \log u \, \cG^{(0)}_S(v,u) + \cG^{(1)}_S(v,u) \right) \,.
\end{equation}

\subparagraph{log(\textit{v}) term}
Let us take the $\log v$ term in equation \eqref{eq:mixedcrossA1}.
We use the fact that $\gamma_{S,0,l}^{(1)} = 0$ to conclude that $\left. \cG_S^{(1)}(v,u)\right|_{v^0 \log v} = 0$.
Projecting out the bilinear part of the crossing equation again, we see that
\begin{equation} \label{eq:mixedcrossA1logv1}
\left. \gamma_\cO^{(1)} \cG^{(0)}_A(u,v)\right|_{u^0  v^{-1}} + \left. \cG^{(1)}_A(u,v)\right|_{u^0  v^{-1}\log v} = 0\, .
\end{equation}
We find this term:
\begin{align}
\left. \cG^{(1)}_A(u,v)\right|_{u^0  v^{-1}\log v} &=
\left. \sum_l a_{SAl}^{(0)}\widetilde{G}_{0,l}(u,v) \right|_{u^0  v^{-1}\log v}  +
\left. \sum_l a_{SAl}^{(0)} \ahat_{SAl}^{(1)} G_{0,l}^{0,0} (u,v) \right|_{u^0  v^{-1}\log v} \nonumber\\
&=  \left. \frac{\gamma_A^{(1)} - \gamma_\cO^{(1)}}{2} \, H^{(0)}_{0,A}(u,v)\right|_{u^0  v^{-1}} + \left. \sum_l a_{SAl}^{(0)} \ahat_{SAl}^{(1)} G_{0,l}^{0,0}(u,v)  \right|_{u^0  v^{-1}\log v}\nonumber \\
&= \left. \frac{\gamma_A^{(1)} - \gamma_\cO^{(1)}}{2} \, \cG^{(0)}_A(u,v)\right|_{u^0  v^{-1}} + \left. \sum_l a_{SAl}^{(0)} \ahat_{SAl}^{(1)} G_{0,l}^{0,0}(u,v)  \right|_{u^0  v^{-1}\log v}\, ,
\end{align}
where we used equation \eqref{eq:1stordermixedconsist}. 

Putting this in the crossing equation \eqref{eq:mixedcrossA1logv1}, we find that
\begin{equation} \label{eq:GAG1logvres}
\frac{\gamma_A^{(1)} + \gamma_\cO^{(1)}}{2} \, \left. \cG^{(0)}_A(u,v)\right|_{u^0  v^{-1}} + \left. \sum_l a_{SAl}^{(0)} \ahat_{SAl}^{(1)}  G_{0,l}^{0,0}(u,v)  \right|_{u^0  v^{-1}\log v}  = 0\, .
\end{equation}
Comparing equations \eqref{eq:tGAtGAcrossres} and \eqref{eq:GAG1logvres}, and using the fact that $\cG_A^{(0)}(u,v)$ and $\widetilde{\cG}_A^{(0)}(u,v)$ have the same bilinear part, we conclude that 
\begin{equation} \label{eq:gamma_adj_is_zero}
\boxed{
\gamma_{A,0,l}^{(1)} = 0
 \, ,
\quad \, l \geqslant 2 \text{ even.}
}
\end{equation}
Specifically note that then \( \alpha_{SAl}\up{1} = \ahat_{SAl}\up{1}\).
Furthermore, matching terms of the form $\frac{u^0}{v}$ in equation \eqref{eq:GAG1logvres}, we find that 
\begin{equation} \label{eq:c1ALcorr}
\boxed{
\alpha_{SAl}\up{1} = 2 \lambda_{SAl}^{(1)} = \left( \gamma_\cO^{(1)} + \gamma_A^{(1)} \right) S_1(l-1) + k_{SA} \,, 
}
\end{equation}
for some constant $k_{SA}$.

\subparagraph{log(\textit{u}) term}
Let us take the $\log u$ term in equation \eqref{eq:mixedcrossA1}.
We use the fact that $\gamma_{A,0,l}^{(1)} =0 $ to conclude that $\left. \cG_A^{(1)}(u,v)\right|_{u^0\log u} = 0$, so that
\begin{equation} \label{eq:mixedcrossA1logu1}
 \frac{ \gamma_\cO^{(1)} +\gamma_A^{(1)}}{2} \cG^{(0)}_S(v,u)\bigg|_{u^{-1}} + \cG^{(1)}_S(v,u)\bigg|_{u^{-1}\log u} = 0\, .
\end{equation}
Recall that the exchanged currents in $\cG_S$ are singlets, which have trivial first-order anomalous dimensions, and which produce conformal blocks with equal external dimensions, so that
\begin{equation} \label{eq:mixedcrossA1logu2}
\left. \cG^{(1)}_S(v,u) \right|_{v^0 u^{-1}\log u} =
\left. \sum_l c_{SSl}\up{0}c_{AAl}\up{0}\left(\lambda_{SSl}\up{1} + \lambda_{AAl}\up{1} \right) G^{0,0}_{0,l}(v,u) \right|_{v^0 u^{-1}\log u} \, .
\end{equation}
The standard analysis then implies that
\begin{equation} 
\lambda_{SSl}\up{1}+\lambda_{AAl}\up{1} = \left( \gamma_\cO^{(1)} +\gamma_A^{(1)}\right) S_1(l-1) + k_S + k_A\, ,
\end{equation}
for some constant $k_A$. 
From this we deduce that
\begin{equation} \label{eq:cAAlcorr}
\boxed{
\lambda^{(1)}_{AAl} =\gamma_A^{(1)}  S_1(l-1) + k_A\, .
 }
\end{equation}

\subsubsection{Bilinear finite support solutions} \label{sec:GNfinsupp}
We would like to discount the possibility of finite support solutions for the bilinear $\gamma_{S,0,l}\up{1}$.
From the analysis in section \ref{sec:2dsinglets}, it is clear that in the singlet correlator $\cG\up{1}(u,v)$, there are no terms of the form $\frac{\log u\log v}{v^k}$ with $k\geq 0$, since there are no $\log J$ terms in the expansion of any anomalous dimensions.

By considering crossing for the mixed correlator $\langle \cO\cO\cO^A\cO^A\rangle$, which relates CFT data for even spin singlet and adjoint operators, the same conclusions hold for the even spin adjoint operators: there are no $\log J$ terms in the expansions of bilinear anomalous dimensions.

For the odd spin adjoint operators, we need to consider the crossing equation for the correlator $\langle \cO^A\cO^A\cO^A\cO^A\rangle$, which is
\begin{equation}
\left. \cG_i\up{1}(u,v)\right|_{\log u \log v} = 
\left. \frac{u}{v}\sum_{j=1}^6 \beta_{ij} \cG_j\up{1}(v,u)\right|_{\log u \log v} \qquad \qquad i=1,\dots,6\,.
\end{equation}
From the above: $\left. \cG_1\up{1}(u,v)\right|_{u^0 \log u \log v} = \left. \cG_2\up{1}(u,v)\right|_{u^0 \log u \log v}= 0$.
Therefore we find that
\begin{align}
\left. \cG_3\up{1}(u,v)\right|_{u^0 \log u \log v} &= 
\frac{u}{v}\sum_{j=3}^6 \beta_{3j} \left. \cG_j\up{1}(v,u)\right|_{u^{-1} \log u \log v}  \\
\left.\cG_i\up{1}(u,v) \right|_{u^0\log u \log v} &=  \frac{u}{v}\sum_{j=3}^6 \beta_{ij} \left. \cG_j\up{1}(v,u)\right|_{u^{-1}\log u \log v}   = 0 \qquad\qquad \text{ for }i\neq 3\, .
\end{align}
We may view the last equation as a set of five linear constraints on four functions $\left. \cG_j\up{1}(u,v)\right|_{v^{-1} \log u \log v} $. 
Generally we expect this to have no non-trivial solutions, and indeed we find that the $\beta_{ij}$ are such that all $\left. \cG_j\up{1}(u,v)\right|_{v^{-1} \log u \log v} $ vanish. 
Hence we conclude that 
\begin{equation}
\left. \cG_3\up{1}(u,v)\right|_{u^0\log u \log v} = 0\, ,
\end{equation}
so that there are no finite support solutions, nor any $\log J$ terms in the expansion of the anomalous dimensions of any of the bilinear currents.

\subsection{Second order} \label{sec:2dSecondOrder}
Once again we need to take into account the dimensional shift due to the theory living in $d=2+\epsilon$. 
Let us consider a fixed TCB of twist $\tau_0$ and expand the various contributions to the twist conformal blocks due to the dimensional shift.
For ease of notation, we shall omit the fact that all functions and derivatives are to be evaluated at $d=d_0=2$ and $\tau=\tau_0$.

Consider the free theory TCB in $d=d_0=2$:
\begin{equation} \label{eq:dimChangeTCBdefn}
H_{\tau_0}\up{0}(u,v) = \sum_l a_{\tau_0,l}\up{0} G\up{d_0}_{\tau_0,l}(u,v)\, .
\end{equation}
In the Gross-Neveu theory in $d=2+\epsilon$, there are corrections:
\begin{align} \label{eq:dimChanges2ndOrder}
a_{\tau_0,l}\up{0} &\rightarrow a_{\tau_0,l}\up{0}\left(1 + \epsilon \widetilde{\alpha}_{\tau_0,l}\up{1}+ \epsilon \alpha_{\tau_0,l}\up{1}  +  \epsilon^2 \widetilde{\alpha}_{\tau_0,l}\up{2}+ \epsilon^2 \alpha_{\tau_0,l}\up{2} +\dots \right)\, , \\
G\up{d_0}_{\tau_0,l}(u,v) 
&\rightarrow \bigg[ 1 +\epsilon\partial_d + \epsilon\left( \gamma_{\tau_0,l}\up{1}+ \zeta \right) \partial_\tau + \epsilon^2\gamma_{\tau_0,l}\up{2} \partial_\tau+ \nonumber\\
&\qquad + \frac{1}{2}\epsilon^2 \left( \left( \gamma_{\tau_0,l}\up{1}+ \zeta \right)^2 \partial_\tau^2 + \left( \gamma_{\tau_0,l}\up{1}+ \zeta \right)\partial_\tau\partial_d+\partial_d^2 \right)+\dots \bigg] G\up{d}_{\tau,l}(u,v)\, ,
\end{align}
where $\widetilde{\alpha}_{\tau_0,l}\up{1}, \widetilde{\alpha}_{\tau_0,l}\up{2}$ are corrections to the free theory due to the dimensional shift, where $\alpha_{\tau_0,l}\up{1}, \alpha_{\tau_0,l}\up{2}$ are corrections due to the departure from the free theory in $d=2+\epsilon$.
Furthermore \( \zeta = \frac{\partial \tau}{\partial d} \) is the spacetime dependence of \( \tau \), i.e. \( \zeta = 1 \) for the bilinear currents and \(\zeta=2\) for the quadrilinear operators.
Gathering the terms in equation \eqref{eq:dimChangeTCBdefn}, we find to order $\epsilon$ the combination with which we are familiar:
\begin{equation}
\underbrace{\sum_l a_{\tau_0,l}\up{0} \left( \widetilde{\alpha}_{\tau_0,l}\up{1} + \zeta \partial_\tau + \partial_d\right) G\up{d}_{\tau,l}(u,v) }_{\text{Free theory correction } \widetilde{\cG}\up{1}(u,v)}
+ \underbrace{\sum_l a_{\tau_0,l}\up{0} \left( \alpha_{\tau_0,l}\up{1} + \gamma\up{1}_{\tau_0,l} \partial_\tau \right) G\up{d}_{\tau,l}(u,v)}_{\text{Gross-Neveu data } \cG\up{1}(u,v)}\, .
\end{equation}
To order $\epsilon^2$, we find the following correction:
\begin{align}
& \quad \underbrace{\sum_l a_{\tau_0,l}\up{0} \left[\widetilde{\alpha}_{\tau_0,l}\up{2}+ \widetilde{\alpha}_{\tau_0,l}\up{1}( \zeta \partial_\tau + \partial_d) + \frac{1}{2}\left(\zeta^2 \partial_\tau^2 + 2 \zeta \partial_\tau\partial_d +\partial_d^2 \right) \right] G\up{d}_{\tau,l}(u,v) }_{\text{Free theory correction }\widetilde{\cG}\up{2}(u,v)} \nonumber\\
& + \underbrace{\sum_l a_{\tau_0,l}\up{0} \left[\alpha_{\tau_0,l}\up{2}  + \gamma\up{2}_{\tau_0,l} \partial_\tau + \alpha_{\tau_0,l}\up{1}\gamma_{\tau_0,l}\up{1}\partial_\tau + \frac{1}{2}\left(\gamma_{\tau_0,l}\up{1}\right)^2 \partial_\tau ^2\right] G\up{d}_{\tau,l}(u,v)}_{\text{Gross-Neveu data } \cG\up{2}(u,v)} \nonumber \\
&  +\underbrace{\sum_l a_{\tau_0,l}\up{0} \left[ \alpha_{\tau_0,l}\up{1}( \zeta \partial_\tau+\partial_d) + \gamma_{\tau_0,l}\up{1} \left(\widetilde{\alpha}_{\tau_0,l}\up{1} +\partial_d + \zeta \partial_\tau \right) \partial_\tau \right] G\up{d}_{\tau,l}(u,v)}_{\text{Cross-term } \nastytwo(u,v)}
\, .
\end{align}
The free theory correction $\widetilde{\cG}\up{2}(u,v)$ can again be calculated by expanding the free theory correlator in $2+\epsilon$ dimensions.
The novelty at this order is the appearance of a cross term $\nastytwo(u,v)$ that combines first-order corrections to the free theory, and first-order departures from the free theory.

Expanding the full crossing relation for the singlets, $v^{\Delta_\cO} \cG(u,v) = u^{\Delta_\cO} \cG(v,u)$ to second order, we find, after much simplification, the relation
\begin{align} \label{eq:cross2ndOrder2d}
 & \gamma_\cO\up{2} \cG\up{0}(u,v)\log v+ \left(\gamma_\cO\up{1} + 1\right) \cG\up{1}(u,v)\log v + \gamma_\cO\up{1}\widetilde{\cG}\up{1}(u,v) \log v  +  \nonumber \\
& \quad + \gamma_\cO\up{1}\left(\frac{1}{2} \gamma_\cO\up{1}+1\right) \cG\up{0}(u,v) \log^2 v + \cG\up{2}(u,v) + \nastytwo(u,v) 
= \frac{u}{v} (u\leftrightarrow v)\, .
\end{align}

Let us consider the $u^0 \log u\log^2 v$ terms on both sides.
Firstly, note that the cross-term could contain divergences of the form $u^0 \log u \log^2 v$.
On the left-hand side of equation \eqref{eq:cross2ndOrder2d}, the relevant term would be 
\begin{equation}
\left. \nastytwo(u,v) \right|_{\log u} = \left. \sum_l a_{\tau_0,l}\up{0} \alpha_{\tau_0,l}\up{1}\partial_\tau G\up{d_0}_{\tau,l}(u,v) \right|_{\log u} 
= \frac{1}{2} \sum_l a_{\tau_0,l}\up{0} \alpha_{\tau_0,l}\up{1} G\up{d_0}_{\tau,l}(u,v)\, ,
\end{equation}
where we used the fact that the bilinear currents satisfy $\gamma_{S,0,l}\up{1}=0$.
Recall that $\alpha_{\tau_0,l}\up{1}$ has been specifically constructed so that this sum is free of $\log^2 v$ divergences, so that $\nastytwo(u,v)$ does not contain a $\log u \log^2 v$ term.
On the right-hand side, the relevant term is due to the quadrilinear operators:
\begin{equation}
\left. C\up{2}(v,u) \right|_{\log u \log^2 v}= \left. \sum_{n,l} \degsum{a_{n,l}\up{0}}\degav{\gamma_{n,l}\up{1}} \partial_\tau^2 G\up{d_0}_{\tau,l}(v,u) \right|_{\log u\log^2v}
 = \left. \sum_{n,l} \frac{1}{4}\degsum{a_{n,l}\up{0}}\degav{\gamma_{n,l}\up{1}} G\up{d_0}_{\tau,l}(v,u) \right|_{\log u}\, .
\end{equation}
However, recall that the $\degav{\gamma_{n,l}\up{1}}$ are precisely of the form guaranteeing that this sum is free of $\log u$ terms.
We therefore see that we can ignore the cross-term in this analysis.

Taking the $u^0 \log u\log^2 v$ term in equation \eqref{eq:cross2ndOrder2d}, we then find the constraint
\begin{equation} \label{eq:2d2ndorderlogulogv2}
\left. \cG \up{2}(u,v)\right|_{u^0\log u \log^2 v} = \frac{u}{v} \left. \cG \up{2}(v,u)\right|_{\frac{\log u \log^2 v}{u}}
\end{equation}
On the left-hand side, $\cG\up{2}(u,v)$ may contain a $u^0 \log u \log^2 v$ term, generated by $\gamma_{S,0,l}\up{2}$, if its expansion in terms of $J_{0,l}^{-2}$ is not constant:
\begin{equation}
\left. \cG\up{2}(u,v)\right|_{\log u\log^2 v} \supseteq \sum_\rho B_{0,\rho}\up{2} H_0\up{\rho}(u,v)\, , \qquad\qquad \gamma_{S,0,l}\up{2} = 2  \sum_\rho B_{0,\rho}\up{2} J_{0,l}^{-2\rho}\, .
\end{equation}
On the right-hand side, the contribution must be as follows:
\begin{equation} \label{eq:2dquadlogulogv2}
\frac{u}{v} \left. \cG \up{2}(v,u)\right|_{\frac{\log u \log^2 v}{u}}  \supseteq 
\left. \frac{u}{v} \sum_{n,l} \frac{1}{8} \degsum{a_{n,l}\up{0}} \degav{(\gamma_{n,l}\up{1})^2} G_{\tau_0(n),l}(v,u) \right|_{\frac{\log u \log^2 v}{u}}\, .
\end{equation}
Note that we had previously found \( \degav{\gamma_{n,l}\up{1}}\); however this does not determine $\degav{(\gamma_{n,l}\up{1})^2}$.
Expanding this sum in terms of the conformal spin:
\begin{equation}
\frac{1}{8}\sum_i \degsum{a_{n,l}\up{0}} \degav{(\gamma_{n,l}\up{1})^2}
= \sum_\rho C_{2n+2}\up{\rho} J_{2n+2,l}^{-2\rho}\, ,
\end{equation}
we get in equation \eqref{eq:2dquadlogulogv2} a contribution
\begin{equation}
\frac{u}{v} \left. \cG \up{2}(v,u)\right|_{\frac{\log u \log^2 v}{u}}  \supseteq  
\left. \frac{u}{v} \sum_{n,\rho} C_{2n+2}\up{\rho}  H_{2n+2}\up{\rho}(v,u)\right|_{\frac{\log u}{u}} \,.
\end{equation}
This term in fact vanishes, which follows from the fact that only the \( H_{2n+2}\up{0}(v,u) \) have \( \frac{1}{u} \) divergences, but do not have any \( \frac{\log u}{u} \) divergences.
Hence
\begin{equation}
\left. \cG \up{2}(u,v)\right|_{u^0\log u \log^2 v} = 0\, ,
\end{equation}
from which it follows that $\gamma_{S,0,l}^{(2)}$ is constant.
Imposing stress-energy tensor conservation then implies that 
\begin{equation}
\boxed{\gamma_{S,0,l}^{(2)} = 0\, .}
\end{equation} 

\subsubsection{Adjoints}
The above analysis applies in the same way to the four-point function of adjoints, and hence we find that the adjoint anomalous dimensions are constant.
Assuming analyticity down to spin $l=2$, there may be a finite support solution for $\gamma_{A,0,1}^{(1)}$, so that our ansatz for the adjoint anomalous dimensions is:
\begin{align}
\gamma_{A,0,l}^{(2)} &= \gamma_{A,0,2}^{(2)} \qquad\qquad\qquad \qquad \, l \geqslant 2  \text{ even} , \nonumber\\
\gamma_{A,0,l}^{(2)} &= \gamma_{A,0,3}^{(2)}\left(1-\delta_{l,1}\right) \qquad\qquad l \geqslant 1  \text{ odd}\, .
\end{align}
These results match the known results in \cite{Giombi:2017rhm}.

\section{The Gross-Neveu-Yukawa model in $d=4-\epsilon$} \label{sec:GNY}
The Gross-Neveu-Yukawa model is a CFT in $d=4-\epsilon$ dimensions providing a perturbation of the free fermion theory.
It has the following action 
\begin{equation} \label{eq:GNYaction}
S = \int \rd^{4-\epsilon}x \left(\psibar^i \slashed{\partial}\psi_i +\frac{1}{2} \left(\partial \sigma\right)^2 + g_1 \psibar^i\psi_i\sigma + g_2 \sigma^4\right) ,
\end{equation}
where $\psi,\psibar$ are conjugate Dirac fermions and $\sigma$ is a scalar field. 
The theory is conformal for a specific value of the pair $(g_1,g_2)$, satisfying $g_1 \sim \sqrt{\epsilon}$ and $g_2 \sim \epsilon$, so that at $\epsilon =0$ it reduces to a 4-dimensional theory of a decoupled free boson and $N_f$ free fermions. 

Our results are for the first order anomalous dimensions of the bilinear currents.
For the adjoint bilinear currents, these are:
\begin{equation}
\gamma_{A,2,l}\up{1} = 2\gamma_\psi^{(1)}\left( 1 - \frac{2}{l(l+1)}\right)\, ,
\end{equation}
for both odd and even spin $l$.

The singlet bilinear currents $J_{\psi,l} \sim \psibar \gamma\partial^{l-1} \psi$ mix with the currents $J_{\sigma,l} \sim \sigma \partial^l \sigma$, and the anomalous dimensions of the resulting primary operators are, for even spin $l$,
\begin{align}
\gamma_{\pm,l}\up{1} = 2\gamma_\psi^{(1)}\left( \frac{N+1}{2} -\frac{1}{l(l+1)} \pm
\frac{ \sqrt{4 + (-4 + 20 N)l(l+1)+ (N-1)^2 l^2(l+1)^2 }}{ 2 l(l+1)} \right)  ,
\end{align}
which were found as the eigenvalues of the following matrix
\begin{equation}
H = 2\gamma_\psi^{(1)}
\begin{pmatrix}
 1- \frac{2}{l(l+1)} & \pm\frac{2\sqrt{N}}{\sqrt{l(l+1)}} \\
\pm\frac{2\sqrt{N}}{\sqrt{l(l+1)}} & N 
\end{pmatrix}.
\end{equation}
This reproduces the results in \cite{Giombi:2017rhm}.

\subsection{Naive attempt at crossing analysis}  \label{sec:GNYnaive}
Let us first look at the four-point correlator of singlets. 
As per the discussion in section \ref{sec:CrossAnalysis}, the crossing equation reads, to first order in $\epsilon$:
\begin{equation}\label{eq:cross1stOrder4d} 
\gamma\up{1}_\cO\log v \, \cG\up{0}(u,v) + \cG\up{1}(u,v) =
\gamma\up{1}_\cO\log u \, \cG\up{0}(u,v) + \frac{u^3}{v^3} \cG\up{1}(v,u)\, .
\end{equation}

We would like to analyse in equation \eqref{eq:cross1stOrder4d}  the power-law divergences in $v$ caused by the bilinears, as in section \ref{sec:CrossFurtherAnalysis}.
As such, we want to compare on both sides the terms $\frac{u}{v^2},\frac{u^2}{v},\frac{u^2}{v^2},\frac{u^2}{v}$.
We define\footnote{This definition differs slightly from that in section \ref{sec:CrossFurtherAnalysis} because $H_2\up{0}(u,v)$ has an integer power-law divergence $v^{-2}$, so that we do not get any power-law divergent terms in $H_2 \up{m}(u,v)$ for $m\geqslant 2$.}
\begin{equation}
H_2^{(m)}( u,v) = \frac{u}{v^2} \widetilde{h}_2^{(m)}(u,v) = \frac{u}{v^2} \left(h_2^{(m)}(u,v) + \cO(u^2) + \cO(v^2)\right) \, , 
\end{equation}

Using the tree-level result and our knowledge of the asymptotic behaviour of TCBs, we deduce that
\begin{equation}
\begin{cases}
h^{(0)}_2(u,v) &= 1-u-v\,, \\
h^{(1)}_2(u,v) &= v\,,\\
h^{(m)}_2(u,v) &= 0 \, ,\qquad\qquad m\geqslant 2\,.
\end{cases}
\end{equation}
Expanding $\gamma_{S,2,l}\up{1}$ and $\ahat_{S,2,l}\up{1}$ as before:
\begin{equation}
\gamma_{S,2,l}\up{1} = 2 \sum_{m=0}^\infty  \frac{B_{2,m}\up{1}}{ J_{2,l}^{2m}}\, , \qquad\qquad 
\ahat_{S,2,l}\up{1} = 2 \sum_{m=0}^\infty A_{2,m}\up{1} \frac{\log J}{J_{2,l}^{2m}} + 2 \sum_{m=0}^\infty \frac{\widetilde{A}_{2,m}\up{1}}{ J_{2,l}^{2m}} \, ,
\end{equation}
we conclude from the crossing equation that 
\begin{equation}
 B_{2,0}\up{1} +  A_{2,0}\up{1}= \gamma_\cO^{(1)}, \qquad B_{2,1}\up{1}= 0 = A_{2,1}\up{1}\, .
\end{equation}
The higher $ B_{2,m}\up{1}$ for $m\geqslant 2$ can be set to zero because the $H_2 \up{m}(u,v)$ develop $\log^2 v$ divergences for $m\geqslant 2$, in such a way that no sum of them is free of $\log^2 v$ terms.
Similarly the higher $A_{2,m}\up{1}$ terms can be set to zero since they would create $\log^3 v$ terms.
The $\widetilde{A}_{2,m}\up{1}$ are then fixed by demanding that the sum 
\begin{equation}
2 A_{2,0}\up{1} H\up{0,\log J}_2 (u,v) + 2 \sum_\rho \widetilde{A}_{2,m} \up{1} H_2\up{\rho}(u,v)
\end{equation}
is free of $\log^2 v$ terms.
As in the 2-dimensional case, we find that this forces the $\widetilde{A}_{2,m}$ to be coefficients in the expansion of the harmonic number $S_1(l)$.

To summarize: we find for the singlets that, in perturbations of the free theory in which the intermediate operators do not change:
\begin{equation}
\gamma_{S,2,l}\up{1} = 2  B_{2,0}\up{1} \, , \qquad\qquad \ahat_{S,2,l} = 2  A_{2,0}\up{1}  S_1(l) + K\, ,
\end{equation}
where $K$ is a constant and $ B_{2,0}\up{1} +  A_{2,0}\up{1}= \gamma_\cO^{(1)}$.

Since the relevant twist conformal blocks are the same for the correlator of four adjoints (up to overall normalizations), the same result can be found for the adjoint currents, in both odd and even spin.
Furthermore, the argument in section \ref{sec:2dMixedCorrs} establishing that the anomalous dimensions of the even singlet and adjoint currents are the same, in facts holds true in any dimension.
Imposing stress-energy conservation, $\gamma_{S,2,l}\up{1} = 0$, would then fix anomalous dimensions and OPE coefficients of the bilinears to be essentially the same as for the Gross-Neveu model in \ref{sec:GN}.

\subsection{Coupling} \label{sec:GNYcoupling}
The results in section \ref{sec:GNYnaive} would hold in a CFT which is a `pure' perturbation of the free fermion theory, i.e. one with no additional operators appearing. 
We are however not aware of any such CFT, and will therefore be interested primarily in the Gross-Neveu-Yukawa model.
The Yukawa interaction in the action \eqref{eq:GNYaction} shows that at order $\epsilon$ one should expect additional operators to appear in the OPE of $\cO \times \cO$.

This leads to two effects.
Firstly, the anomalous dimensions and OPE coefficients in the previous section may acquire corrections. 
Secondly, there will be a another set of bilinear currents $J_{\sigma,l} \sim \sigma \partial^l \sigma$ of twist $2$, which mix with the bilinear currents $J_{\psi,l}$. 

Let us therefore do a more conservative analysis than that in section \ref{sec:GNYnaive}, and focus on making sure the crossing relation contains no terms of the form $\log u \log^m v$ for $m\geqslant 2$.
From the free theory result for the twist conformal block $H_2\up{0}(u,v)$ of the bilinears, we can calculate all $H_2\up{m}(u,v)$.
We find that $H_2\up{m}(u,v)$ contains a term of the form $h\up{m}(v) \log^2 v$ for $m \geqslant 2$, with $h\up{m}(v) \sim v^{m-2}$ for small $v$.
Therefore, we exclude any terms of the form $J^{-2m}$ with $m\geqslant 2$ in the expansion of $\gamma_{S,2,l}\up{1}$, and in fact we claim that there can be no terms of the form $\frac{\log^k J}{J^{2m}}$ in its expansion, so that it is of the form
\begin{equation}
\gamma_{S,2,l}\up{1} = A\left(1-\frac{B}{l(l+1)}\right)\, .
\end{equation}
To see why this holds, consider the case $k=0$. 
From the above, we see that any terms $\frac{\log J}{J^{2m}}$ with $m\geqslant 2$ will contain divergences of the form $\log^3 v$, and must therefore be discarded.
We focus on the $\log u \log v$ part of the crossing equation:
\begin{equation}
\left. \cG\up{1}(u,v)\right|_{\log u \log v} = \frac{u^3}{v^3} \left. \cG\up{1}(v,u)\right|_{\log u \log v}\, , 
\end{equation}
and specifically terms of the form $u^1 v^{-2}, u^1 v^{-1}, u^2v^{-2}, u^2v^{-1}$.
These terms arise only from bilinears with infinite support on the spin, and are taken onto themselves under crossing symmetry.
Specifically additional operators appearing in the OPE cannot give a $\log u \log v$ term, so we may ignore them.
An analysis of the precise divergences then shows that no term of the form $\frac{\log J}{J^2}$ can appear.
To show there is no term of the form $\gamma_{S,2,l}\up{1} = 2 \beta S_1(l)+\dots$, we analyse all terms in the crossing equation of the form \( \frac{u^n}{v^m} \log u \log v\), where \(m>0\).
These terms can only arise from harmonic number behaviour of anomalous dimensions. 
Note that the quadrilinear operators of \( \tau_0(n) = 6+2n \) cannot have a \( \log J\) behaviour in their anomalous dimensions, since under crossing this would create a correction to the identity operator.
We therefore make the ansatz
\begin{align}
\gamma_{S,2,l}\up{1} = 2\beta S_1(l) + \dots \, , \qquad\qquad
\gamma_{S,6+2n,l}\up{1} = 2\beta_n \frac{S_1(l+n+2)}{J_{6+2n,l}^2} + \dots\, .
\end{align}
Matching the \( \log u \log v \) part of the crossing equation then yields:
\begin{equation}
\beta H^{(0),harm}_2(u,v)\bigg|_{\log v} + \sum_{n=0}^\infty \beta_n H^{(1),harm}_{6+2n}(u,v)\bigg|_{\log v} \stackrel{\boldsymbol{\cdot}}{=} \beta \frac{u^3}{v^3} H^{(0),harm}_2(v,u)\bigg|_{\log u} \, ,
\end{equation}
where by \(\stackrel{\boldsymbol{\cdot}}{=}\) we mean that only power-divergent terms in \( v \) are matched, and where \(H_{\tau_0}^{(m),harm}(u,v) \) is the \(m\)-th twist conformal block with a harmonic number insertion:
\begin{equation}
 H_{\tau_0}^{(m),harm}(u,v) \equiv \sum_l a\up{0}_{\tau_0,l} J_{\tau_0,l}^{-2m} S_1\left( l+ \frac{\tau_0}{2}-1\right) G_{\tau_0,l}(u,v)\, .
\end{equation}
We find that this equation has only the trivial solution \( \beta=\beta_n=0\).

We therefore conclude that no terms of the form $\frac{\log J}{J^{2m}}$ appear in the expansions of the anomalous dimensions of any of the intermediate operators.
Terms with higher powers of $\log J$ are similarly excluded.

Note that the argument relies on the assumption that the CFT data is analytic in the spin down to spin \(l=2\), so that finite support solutions for the bilinear operators could be excluded.

\subsubsection{Finite support solutions} \label{sec:GNYfinsupp}
Like for the 2d Gross-Neveu model, we would like to discount the possibility of finite support solutions for the bilinear $\gamma_{A,2,l}\up{1}$.
The discussion is isomorphic to that in section \ref{sec:GNfinsupp}, with as its only difference that some of the powers of $u$ and $v$ change.
Specifically, consideration of terms of the form $u^k \log u \log v$ for $k=0,1,2$ show that there are no terms of the form $v^{-k} \log u \log v$ for $k=1,2,3$, i.e. there are no enhanced divergences proportional to $\log u \log v$.
We therefore conclude that there are no finite support solutions for the anomalous dimensions of the bilinear currents.

\subsubsection{Mixing of singlet currents} \label{sec:GNYmixing}
In the Gross-Neveu-Yukawa model, there are two sets of singlet operators of twist $\tau = d-2$: the currents $J_{\psi,l}$, and the currents $J_{\sigma,l} \sim \sigma \partial^l \sigma$.
These currents mix due to a coupling between $\psi$ and $\sigma$, so that they are no longer eigenstates of the Hamiltonian/dilatation operator.
Instead, the eigenstates of the Hamiltonian will be
\begin{align}
\Sigma_{-,l} &= A_- \, J_{\psi,l} + B_- \, J_{\sigma,l} \, , \\
\Sigma_{+,l} &= A_+ \, J_{\psi,l} + B_+ \, J_{\sigma,l} \, , 
\end{align}
whose eigenvalues $\Delta_{\pm,l} = d-2+ \gamma_{\pm,l}$ under the Hamiltonian are their scaling dimensions.

Phrased differently: in the basis $\{J_{\psi,l},J_{\sigma,l}\}$ the Hamiltonian does not act diagonally.
Let us represent it as a matrix in this basis:
\begin{equation} \label{eq:mixhamildef}
H = \begin{pmatrix}
\langle J_{\psi,l} | \hat{H} | J_{\psi,l}  \rangle  
& \langle J_{\sigma,l} | \hat{H} | J_{\psi,l}  \rangle \\ 
\langle J_{\psi,l} | \hat{H} | J_{\sigma,l}  \rangle 
& \langle J_{\sigma,l} | \hat{H} | J_{\sigma,l}  \rangle 
\end{pmatrix}
=  H_0 + \epsilon H_\epsilon + \dots =
H_0 + \epsilon \begin{pmatrix}
a_l & c_l \\ c_l & d_l
\end{pmatrix}
+\dots \, ,
\end{equation}
which is symmetric since the Hamiltonian is self-adjoint.
Our goal shall be to find \(a_l,c_l,d_l\), and deduce from it the anomalous dimensions \( \gamma_{\pm,l}\up{1}\).

Since $\cO$ only couples to $\sigma$ at order $\epsilon$, we find that 
\begin{equation}
\langle \cO \cO \Sigma_\pm \rangle = A_\pm \langle \cO \cO J_\psi \rangle + O(\epsilon) \, ,
\end{equation}
so that, to zeroth order in $\epsilon$,
\begin{equation}
A_\pm = \frac{\langle \cO \cO \Sigma_\pm \rangle}{\langle \cO \cO J_\psi \rangle} \, .
\end{equation}
The $\gamma_{S,2,l}^{(1)}$ calculated in sections \ref{sec:GNYnaive} and \ref{sec:GNYcoupling} is then defined as the average over both eigenstates:
\begin{equation}
\gamma_{S,2,l}^{(1)} \langle \cO \cO J_\psi \rangle^2  = \sum_i \langle \cO \cO \Sigma_\pm \rangle^2 \gamma_{\pm,l}\up{1} \,.
\end{equation}
However since the $\Sigma_\pm$ are precisely the vectors that diagonalize the symmetric matrix $H_\epsilon$, the above is simply an entry of $H_\epsilon$
\begin{equation}
\gamma_{S,2,l}^{(1)} = \sum_i A_\pm^2 \gamma_{\pm,l}\up{1} = a_l \, .
\end{equation}

Furthermore, recall that at $\epsilon= 0$, $\sigma$ is a free boson, so that the analysis from \cite{Alday:2016jfr} applies.
It shows that the first order correction $d_l$ to $\langle J_{\sigma,l} | \hat{H} | J_{\sigma,l} \rangle$, is of the form 
\begin{equation}
d_l = 2 \gamma^{(1)}_\sigma\, .
\end{equation}
Then $H_\epsilon$ is of the form
\begin{equation}
H_\epsilon = \begin{pmatrix}
A\left(1- \frac{B}{l(l+1)}\right)
& C_l \\ 
C_l & 2 \gamma^{(1)}_\sigma
\end{pmatrix}.
\end{equation}
The twist spectrum additivity property of \cite{Komargodski:2012ek,Fitzpatrick:2012yx} implies that:
\begin{equation}
\lim_{l\rightarrow \infty} \{\gamma_{-,l}, \gamma_{+,l} \} = \left\{2\gamma_\psi^{(1)},2\gamma_\sigma^{(1)} \right\}\, .
\end{equation}
Assuming that $\gamma^{(1)}_\psi \neq \gamma^{(1)}_\sigma$, this imposes the constraint $A = 2\gamma^{(1)}_\psi$, and that $\lim_{l\rightarrow \infty} C_l = 0$, so that
\begin{equation}
H_\epsilon = \begin{pmatrix}
2\gamma^{(1)}_\psi \left(1 -\frac{B}{l(l+1)}\right)
& C_l \\ 
C_l & 2 \gamma^{(1)}_\sigma
\end{pmatrix}
\end{equation}
with $\lim_{l\rightarrow \infty} C_l = 0$.

In order to proceed we need to find the off-diagonal terms $C_l = \langle J_{\psi,l} | H | J_{\sigma,l}\rangle$.
This naturally leads us to consider the mixed correlator $\langle \cO\cO\sigma\sigma \rangle$, which maps to $\langle  \sigma\cO\cO \sigma\rangle$ under crossing:
\begin{equation}
\qquad\cG_{ \sigma \cO\cO \sigma}(u,v) = \frac{u^2}{v^3}\cG_{ \cO \cO \sigma\sigma}(u,v)\, .
\end{equation}
Using Wick contractions, we can calculate the free theory results:
\begin{equation}
\cG_{ \cO \cO \sigma\sigma}\up{0}(u,v) = 1\, , \qquad\cG\up{0}_{ \sigma \cO \cO \sigma}(u,v) = \frac{u^2}{v^3}\,.
\end{equation}
We will also need the free theory result for the related correlator \( \langle  \cO\sigma\cO \sigma\rangle \), which satisfies
\begin{equation}
\cG_{ \cO\sigma\cO \sigma}\up{0} (u,v) = u^2\, , \qquad \cG_{ \cO\sigma\cO \sigma}(u,v) = \frac{u^2}{v^2}\cG_{ \cO\sigma\cO \sigma}(v,u)\,.
\end{equation}
In the ``direct channel'' expansion of $\langle\cO\cO\sigma\sigma\rangle$, we simply see the contribution of the identity. 
In particular, we see no contribution from the bilinear currents $J_{\psi,l} \sim \psibar \gamma \partial^{l-1}\psi$ and $J_{\sigma,l} \sim \sigma \partial^l \sigma$. 
This is expected, since at tree-level these currents only couple to one of $\cO$ and $\sigma$, and thus cannot function as intermediate states.
However, at tree-level the Hamiltonian is diagonal in the space spanned by $J_{\psi,l},J_{\sigma,l}$; therefore, with the benefit of foresight, let us define the rotated states $\Sigma_{\pm,l}$ by
\begin{align}
\Sigma_{-,l} &= \cos \theta_l \, J_{\psi,l} - \sin \theta_l \, J_{\sigma,l} \,, \\
\Sigma_{+,l} &= \sin \theta_l \, J_{\psi,l}  + \cos \theta_l \, J_{\sigma,l}\, .
\end{align}
As \( \epsilon \) turns on, we shall choose the \( \theta_l \) so that these are the eigenstates of \( \hat{H}\).
Note the (potential) $l$-dependence of the angle $\theta_l$.\footnote{We assume here that these bases of states can be related by a rotation.
There is also the possibility of the transformation being a reflection; however this will not affect the discussion in this section.}

The Hamiltonian still acts diagonally on these states, and thus we may view them as two different intermediate states propagating in the  direct channel.
They give cancelling contributions of 
\begin{equation}
\pm \tilde{H}^{(0)}_2(u,v) = \pm\, \sum_{l\in 2\mathbb{N}} \sin\theta_l\cos\theta_l  c^{(0)}_{\cO\cO J_{\psi,l}}c^{(0)}_{\sigma\sigma J_{\sigma,l}}G_{2,l}(u,v)
\end{equation} 
to the tree-level result.

In the Gross-Neveu-Yukawa model there will be some $\theta_l$ for which $\Sigma_{\pm,l}$ are the eigenstates of the Hamiltonian.
These states may acquire different anomalous dimensions, and as such, the contribution above will generally change to include an order $\epsilon$ term
\begin{equation}
\sum_{l\in 2\mathbb{N}}  \frac{1}{2}\sin\theta_l\cos\theta_l \left(\gamma^{(1)}_{+,l}-\gamma^{(1)}_{-,l}\right)  c^{(0)}_{\cO\cO J_{\psi,l}}c^{(0)}_{\sigma\sigma J_{\sigma,l}} \partial_\tau\big|_{\tau=2} G_{\tau,l}(u,v)\, .
\end{equation} 
We would like to consider this sum in the language of twist conformal blocks; as such it would be useful to calculate 
\begin{equation}
H^{(0)}_2(u,v)\equiv \sum_{l\in 2\mathbb{N}} c^{(0)}_{\cO\cO J_{\psi,l}}c^{(0)}_{\sigma\sigma J_{\sigma,l}}G_{2,l}(u,v) \, .
\end{equation}
An interesting point to note here is that the OPE coefficients $ c^{(0)}_{\cO\cO J_{\psi,l}}$ and $c^{(0)}_{\sigma\sigma J_{\sigma,l}}$ in the free fermion and boson theories are related:
\begin{equation} \label{eq:bosferoperelation}
c^{(0)}_{\cO\cO J_{\psi,l}} = \sqrt{l(l+1)} c^{(0)}_{\sigma\sigma J_{\sigma,l}}\, 
\end{equation}
which is the reason that in the free fermion theory correlator $\langle\cO\cO\cO\cO\rangle$, the twist conformal block $H^{(1)}_{2,F}$ is the $H^{(0)}_{2,B}$ of the free boson correlator $\langle\sigma\sigma\sigma\sigma\rangle$.
Phrased differently, the free fermion TCB is simply the result of applying the Casimir operator $\mathcal{C}$ to the free boson TCB.
Therefore equation \eqref{eq:bosferoperelation} implies that
\begin{equation} \label{eq:mixReltoBos}
H^{(0)}_2(u,v) = H_{2,B}^{\left(-\frac{1}{2}\right)}(u,v) = H_{2,F}^{\left(\frac{1}{2}\right)}(u,v) \, ,
\end{equation}
where $H^{(m)}_{2,B}$ and $H^{(m)}_{2,F}$ are the free boson, respectively free fermion, twist conformal blocks of twist $2$.
\\
\\
Using the method in \cite{Alday:2015eya}, we have been able to evaluate the small-$u$ limit:\footnote{We have been able to compute it to a finite order in $v$.
Theoretically we can calculate to any order in $v$, but computation time prevents us from doing this to higher order.
Regardless, the discussion that follows only relies on the structure of this function, not its precise coefficients.
}
\begin{equation}
H^{(0)}_2(u,v) = \frac{\pi}{4} \frac{u}{v^{3/2}} \left(1-v-\frac{v^2}{2} -\frac{7 v^3}{18}-\frac{203 v^4}{600} + \dots \right) + \cO(u^2)\, .
\end{equation}
Let us investigate the implications of this.
The twist conformal blocks have the following behaviour:
\begin{equation}
H^{(\rho)}_2(u,v) \equiv  \sum_{l\in 2\mathbb{N}} J_{2,l}^{-2\rho} c^{(0)}_{\cO\cO J_{\psi,l}}c^{(0)}_{\sigma\sigma J_{\sigma,l}}G_{2,l}(u,v) \sim \frac{u}{v^{\frac{3}{2}-\rho}}(1+\cO(v) ) + \cO(u^2) \, .
\end{equation}
Expanding, as usual:
\begin{equation}
 \frac{1}{2}\sin\theta_l\cos\theta_l \left(\gamma^{(1)}_{+,l}-\gamma^{(1)}_{-,l}\right) = \sum_\rho B_\rho J_{2,l}^{-2\rho} \, ,
\end{equation}
we are interested in the sum
\begin{equation}
\sum_\rho B_\rho H^{(\rho)}_2(u,v) \subset \cG^{(1)}_{\cO\cO\sigma\sigma}(u,v) \, ,
\end{equation}
where the sum over $\rho$ now includes half-integers to allow for the possibility of odd powers of $J^{-1}_{2,l}$ in the expansion of $\sin\theta_l\cos\theta_l \left(\gamma^{(1)}_{+,l}-\gamma^{(1)}_{-,l}\right)$.

From equation \eqref{eq:mixReltoBos} and the earlier discussion in section \ref{sec:GNYcoupling}, one deduces that $H^{(\rho)}_2(u,v)$ contains $\log^2 v$ terms for $\rho = \frac{3}{2},\frac{5}{2},\frac{7}{2},\dots$, in such a way that no linear combination of them is free of $\log^2 v$ terms.
Thus 
\begin{equation}
B_{\rho} = 0\, , \qquad  \qquad \text{for } \rho = \frac{3}{2},\frac{5}{2},\frac{7}{2},\frac{9}{2},\dots \, . 
\end{equation}
Furthermore, we shall show that $B_{\rho}=0$ for all $\rho=0,1,2,3\dots$.
To see this, we use the crossing equation
\begin{equation}
\cG_{\sigma\cO\cO\sigma}(u,v) = \frac{u^2}{v^3} \cG_{\cO\cO\sigma\sigma}(v,u)\supset \epsilon \frac{u^2}{v^3}\sum_{\rho=0,\frac{1}{2},1,2,3,\dots} B_\rho H^{(\rho)}_2(v,u) \log v \, .
\end{equation}
Note that 
\begin{equation}
\frac{u^2}{v^3} H^{(\rho)}_2(v,u) \sim \frac{u^{\rho+\frac{1}{2}}}{v^2}\, .
\end{equation}
Therefore \( \cG_{\sigma\cO\cO\sigma}(u,v)  \) contains a term of the form \( u^{\rho+\frac{1}{2}}v^{-2} \log v \).
Then \( \cG_{\cO\sigma\cO\sigma}(u,v) \) will necessarily contain a term of the form \( u^{\rho+\frac{1}{2}}v^{-2+k} \log v \) for some integer \( k \geqslant 0 \). 
The crossing equation for \( \cG_{\cO\sigma\cO\sigma}(u,v) \) implies that then also
\begin{equation} \label{eq:OSigmaOSigmaCorr}
\cG_{\cO\sigma\cO\sigma}(u,v) \supseteq \frac{u^k}{v^{\frac{3}{2}-\rho}} \log u\, .
\end{equation}
This term must arise from a non-zero anomalous dimension for any operator that is already present in the free theory.\footnote{In the definition of present we also allow mixed operators such as \( \Sigma_{\pm,l} \) that give cancelling contributions to the tree-level result, since they can yield terms of the form \( \epsilon \log u \) in \(\cG\up{1}(u,v) \).}
It is easy to see that in the free theory, \( \langle \cO \sigma \Sigma_{\pm,l} \rangle = 0 \), so that \( k \geqslant 2 \).
From the tree-level result \( \cG\up{0}_{\cO\sigma\cO\sigma}(u,v) = u^2 \), we conclude that there are operators of twists \( 4, 6, \dots \).
Decomposing the tree-level result, we see that the intermediate operators of odd and even spin give similar, almost cancelling contributions that have a power-law divergence of the form \( v^{-2} \):
\begin{align}
\cG\up{0}_{\cO\sigma\cO\sigma}(u,v) &= \cG^{(0),even}_{\cO\sigma\cO\sigma}(u,v) + \cG^{(0),odd}_{\cO\sigma\cO\sigma}(u,v) \nonumber\\
&= u^2\left(1 + \left(\frac{1}{4v^2}+\frac{1}{4v}- \frac{1}{2} \right)\right) - u^2 \left(\frac{1}{4v^2}+\frac{1}{4v}- \frac{1}{2} \right) + \cO(u^3)\, .
\end{align}
We have shown this property here for the twist 4 operators, but due to the special form of the 4d conformal blocks, it in fact holds for all higher twist operators as well.
If the even and odd spin intermediate operators acquire different anomalous dimensions, then there can be \( \log u \) corrections to the correlator of the form \( u^n v^m \) for integer \( n\geqslant 2, m\geqslant -2 \).
Specifically note that integrality of the powers of \( v \) forces, by virtue of equation \eqref{eq:OSigmaOSigmaCorr}, that \( \frac{3}{2} - \rho \in \mathbb{Z} \), so that all integer \( \rho \) are excluded.
Therefore \( \rho=\frac{1}{2} \) is the only remaining possibility, from which it follows that
\begin{equation}
\boxed{
 \frac{1}{2}\sin\theta_l\cos\theta_l \left(\gamma^{(1)}_{+,l}-\gamma^{(1)}_{-,l}\right) = \frac{B_{1/2}}{J_{2,l}} \, .
 }
\end{equation}
Looking at the diagonalization of the Hamiltonian $H$, this is precisely the off-diagonal entry! 
That is, we have found that, in the basis $\{J_{\psi,l}, J_{\sigma,l} \}$, the order $\epsilon$ correction to the Hamiltonian takes the form
\begin{equation}
H_\epsilon= \begin{pmatrix}
2\gamma_\psi\up{1} \left( 1- \frac{B}{l(l+1)}\right) & \frac{2B_{1/2}}{\sqrt{l(l+1)}} \\
\frac{2B_{1/2}}{\sqrt{l(l+1)}} & 2\gamma_\sigma^{(1)} \, .
\end{pmatrix}
\end{equation}

To fix the constants $B,B_{1/2}$, we factor out $2\gamma_\psi\up{1}$ to get the matrix
\begin{equation}
\tilde{H} = \begin{pmatrix}
 1- \frac{B}{J^2} & \frac{C}{J} \\
\frac{C}{J} & \omega \, ,
\end{pmatrix}
\end{equation}
where $\omega = \gamma_\sigma^{(1)}/ \gamma_\psi^{(1)}$.

We now use a central charge argument to fix $B$ and $C$ in terms of $\omega$ and $N$.
Recall that a 4d free theory of $N_B$ free scalar fields and $N_f$ free Dirac fermions has a central charge
\begin{equation}
C_T = c_B + c_F = \frac{4}{3} N_B + 8 N_f = \frac{4}{3} N_B + 2 N \, ,
\end{equation}
where $N = N_f \text{tr}\mathds{1} = 4 N_f$. 	
Furthermore, recall the relation between the central charge $c_T$ and OPE coefficients with the stress-energy tensor: for any operator $\cO$ in a $d$-dimensional theory with a (unique) stress-tensor $T$,  the following relation holds \cite{Komargodski:2012ek}:
\begin{equation}
a_{\cO\cO T} \equiv \frac{C_{\cO\cO T}^2}{C_{TT}C_{\cO\cO}^2} = \frac{1}{c_T} \frac{d^2}{(d-1)^2} \Delta_{\cO}^2\, ,
\end{equation}
where the $C_{\bullet\bullet}$ are two-point normalizations.
Thus, in the theory of free fermions and free bosons:
\begin{equation}\label{eq:freeTheoryCentralChargeRels}
\frac{1}{c_F} \frac{d^2}{(d-1)^2} \Delta_{\psi^2}^2= \frac{C_{\psi^2\psi^2 T_F}^2}{C_{T_F T_F}C_{\psi^2\psi^2}^2}\, , \qquad \qquad 
\frac{1}{c_B} \frac{d^2}{(d-1)^2} \Delta_{\sigma}^2= \frac{C_{\sigma\sigma T_B}^2}{C_{T_B T_B}C_{\sigma\sigma}^2}\, .
\end{equation}
In the GNY model at $\epsilon=0$, there are two decoupled free theories, with central charges $c_B = \frac{4}{3}\, , \, c_F = 2N$ and two separate stress-energy tensors $T_B = J_{\sigma,2},\, T_F = J_{\psi,2}$ satisfying equation \eqref{eq:freeTheoryCentralChargeRels}.
As we turn on $\epsilon$, there will be a unique stress tensor $T = \Sigma_{-,2}$, and a single central charge $c_T$ satisfying
\begin{equation}
\lim_{\epsilon\rightarrow 0} c_T = c_B + c_F = \frac{4}{3} + 2 N , 
\end{equation}
and 
\begin{align}
\frac{1}{c_T}\frac{d^2}{(d-1)^2} \Delta_{\psi^2}^2 &= a_{\psi^2\psi^2 T}^2
= \frac{C_{\psi^2\psi^2 T}^2}{C_{T T}C_{\psi^2\psi^2}^2} \label{eq:fermionOPEConstraint} \, ,\\
\frac{1}{c_T}\frac{d^2}{(d-1)^2} \Delta_{\sigma}^2 &= a_{\sigma\sigma T}^2
= \frac{C_{\sigma\sigma T}^2}{C_{T T}C_{\sigma\sigma}^2}\, .
\end{align}
We find that the last two equations give the same constraint, so we shall only use equation \eqref{eq:fermionOPEConstraint}.
Furthermore, we shall only be interested in the free theory limit $\epsilon \rightarrow 0$.

Assuming proper normalization of $J_{\psi,l}$ and $J_{\sigma,l}$ (i.e. $C_{J_{\bullet,l}J_{\bullet,l}}=1$), we can write
\begin{equation}
\begin{pmatrix}
\Sigma_{-,l}\\ \Sigma_{+,l}
\end{pmatrix}
=
\begin{pmatrix}
\alpha_l&\beta_l \\
\gamma_l&\delta_l
\end{pmatrix}
\begin{pmatrix}
J_{\psi,l}\\ J_{\sigma,l}
\end{pmatrix},
\end{equation}
where both the columns and rows of the matrix form orthonormal vectors.\footnote{We are explicitly allowing for a reflection as well as a rotation here.}
Note that $\alpha_l,\beta_l,\gamma_l,\delta_l$ are functions of $B,C,\omega$ and $l$.
Considering only $\epsilon^0$ terms, we find that
\begin{equation}
\langle \psi^2 \psi^2 \Sigma_{-,2} \rangle^2 = \alpha_2^2 \langle \psi^2\psi^2 J_{\psi,2}\rangle^2
\end{equation}
and that
\begin{equation}
\langle \Sigma_{-,2} \Sigma_{-,2} \rangle = \alpha_2^2  \langle  J_{\psi,2}  J_{\psi,2}\rangle^2 + \beta_2^2  \langle  J_{\sigma,2}  J_{\sigma,2}\rangle^2\, = \alpha_2^2+\beta_2^2 .
\end{equation}

Then equation \eqref{eq:fermionOPEConstraint} yields
\begin{align}
\frac{1}{c_F+c_B}\frac{d^2}{(d-1)^2} \Delta_{\psi^2}^2   
= \frac{\alpha_2^2 C_{\psi^2\psi^2 T_F}^2}{C_{\psi^2\psi^2}^2(\alpha_2^2  + \beta_2^2) }
=  \frac{C_{\psi^2\psi^2 T_F}^2}{ C_{\psi^2\psi^2}^2} \frac{1}{1+ \left(\frac{\beta_2}{ \alpha_2}\right)^2 }
= \frac{1}{c_F}\frac{d^2}{(d-1)^2} \frac{1}{1+x^2} \Delta_{\psi^2}^2   \, ,
\end{align}
where $x(B,C,\omega)\equiv\frac{\beta_2}{ \alpha_2}$. 
Thus we get a constraint on $B,C$: $x(B,C,\omega)^2 = \frac{c_B}{c_F}$.

Furthermore, since $\Sigma_{-,2}$ is the stress tensor, we also have the constraint $\gamma_{-,2}(B,C,\omega) = 0$.
These two constraints are independent and yield the following solutions for $B,C$:
\begin{equation} \label{eq:4dGNYBCsol}
B = 6 - 4\frac{\omega}{N}, \qquad\qquad C= \pm 2\frac{\omega}{\sqrt{N}}\,. 
\end{equation}
Unfortunately we are not able to fix the sign on $C$; however note that the anomalous dimensions are not sensitive to the sign of $C$.

Finally, we are unable to fix \( \omega \) from a bootstrap argument and instead get the value \( \omega = N \) from the literature, so that
\begin{equation} \label{eq:4dSingBilAnom}
\langle J_{\psi,l} | \hat{H} | J_{\psi,l} \rangle= 2\gamma_\psi\up{1}\left(1 - \frac{2}{l(l+1)} \right)\, .
\end{equation}
The argument from section \ref{sec:2dMixedCorrs} that the even spin singlet and even spin adjoint currents have the same anomalous dimensions\footnote{To be more precise: the argument that \(\langle J_{\psi,l} | \hat{H} | J_{\psi,l} \rangle = \langle J_{\psi,l}^A | \hat{H} | J_{\psi,l}^A \rangle \) for even spin \( l \geqslant 2 \).}, is in fact independent of the dimension of the space, so that the adjoint currents of even spin have an anomalous dimension as in equation \eqref{eq:4dSingBilAnom}.
Furthermore, the conservation of the global symmetry current implies that the anomalous dimensions of the odd spin adjoint currents take the same form, so that for both odd and even \( l \):
\begin{equation}
\boxed{ \gamma_{A,2,l}\up{1} = 2 \gamma_\psi\up{1} \left(  1- \frac{2}{l(l+1)}\right)} \qquad\qquad l\geqslant 1 \, .
\end{equation}

Using the minus sign for $C$ in equation \eqref{eq:4dGNYBCsol}, we reproduce the known result for the singlet currents \cite{Giombi:2017rhm}:
\begin{equation}
\boxed{ H = 2\gamma_\psi^{(1)}
\begin{pmatrix}
 1- \frac{2}{l(l+1)} & -\frac{2\sqrt{N}}{\sqrt{l(l+1)}} \\
-\frac{2\sqrt{N}}{\sqrt{l(l+1)}} & N 
\end{pmatrix} ,
}
\end{equation}
which gives the anomalous dimensions as
\begin{equation}
\gamma\up{1}_{\pm,l}= 2\gamma_\psi^{(1)}\left( \frac{N+1}{2} -\frac{1}{l(l+1)} \pm
\frac{ \sqrt{4 + (-4 + 20 N)l(l+1)+ (N-1)^2 l^2(l+1)^2 }}{ 2 l(l+1)} \right) \, .
\end{equation}

\section{Discussion} \label{sec:discussion}
In this paper we have used crossing symmetry to constrain fermionic CFTs that weakly break higher spin symmetry through the study of the analytic properties of the twist conformal blocks occurring in the four-point correlators of composite operators.
Novel to the use of composite operators is that, in contrast to the previous study of correlators of fundamental scalar fields \cite{Alday:2016jfr}, quadrilinear operators appear as intermediate states already in the free theory correlator.
Their CFT data mixes under crossing with that of the bilinear operators, making it harder to isolate the CFT data of the bilinear operators.
As has been found in previous work (see e.g. \cite{Kos:2014bka}), the bootstrap gains tremendously in power when several different correlators are studied simultaneously, demonstrated in our paper by the study of mixed correlators such as $\langle \cO \cO^A \cO^A \cO \rangle$ in both models, and the study of $\langle \cO\cO\sigma\sigma \rangle$ in the Gross-Neveu-Yukawa model.
Our method reproduces known results for the anomalous dimensions of bilinear currents \cite{Giombi:2017rhm}, produces some new results for bilinear OPE coefficients, and finds CFT data of the quadrilinear operators.
Furthermore it finds a solution in \(2\) dimensions for a potential fermionic CFT in which the fundamental field \( \psi\) is not in the spectrum.
Some extensions of our work are clear.

Compared with the analysis of correlators of fundamental scalar fields \cite{Alday:2016jfr}, our method suffers from the obvious drawback that there are a larger number of intermediate operators, making it harder to isolate contributions of any particular intermediate operator to the four-point correlator.
It would be interesting to extend the method of the large spin bootstrap to include correlators of non-scalar operators to facilitate the study of the correlator of four fundamental fermion fields.
In \cite{Cuomo:2017wme} a formalism to study such correlators in four dimensions is established; it would be interesting to try to study the Gross-Neveu-Yukawa model using these methods.

The Gross-Neveu model can be defined in any dimension \( 2 < d < 4 \) through the large \( N \) expansion, and results for CFT data are known perturbatively in \( 1/N \) \cite{Giombi:2017rhm}.
They have the interesting property that they are essentially identical to corrections in the bosonic critical large \( N \) model; it would be interesting to apply the method of the large spin bootstrap to try to understand this.
Unfortunately the computations become a lot more complicated in the large \( N \) model; see for example the increased complexity in the discussion of the bosonic critical large \( N \) model in \cite{Alday:2016jfr}.

Finally we have had to deal in an ad-hoc manner with the dimensional shift arising from the non-integer dimension of spacetime.
The difficulty of dealing with this increases significantly with each order in \(\epsilon\).
The development of a systematic method to deal with these issues should simplify calculations and hopefully allow an (easier) analysis of higher-order corrections.
Of particular interest would be an application of such methods to the Wilson-Fisher model in \( d=4-\epsilon \), where the extension of the large spin bootstrap to new orders in \( \epsilon \) is hampered by the issues of dealing with the non-integer dimension of the spacetime in which the theory lives.

\section*{Acknowledgements}
I would like to thank Luis F. Alday, Tomasz {\L}ukowski, Johan Henriksson and Simone Giombi for helpful discussions.
I also thank the organisers of the Pollica Summer Workshop 2017, where part of this work was presented, for their hospitality.
During completion of this work, the author was supported by ERC STG grant 306260, and by an EPSRC studentship.

\appendix

\section{Summary of results} \label{app:results}
The bilinear operators of spin \(l\) are of the form \( J_l \sim \psibar \gamma \partial^{l-1}\psi  \), and occur both in the singlet and adjoint representation of the global \( U(N_f) \) symmetry.
We shall refer to them as \( J_{S,l} \) and \( J_{A,l} \) respectively.
The quadrilinear operators are operators built of four fundamental fields, with a number of derivatives acting on them.
For example, twist 4 quadrilinear operators are of the form \( \psibar \partial^{l_1} \psi \partial^{l_2} \psibar \partial^{l_3} \), where \( l = l_1 + l_2 + l_3 \) is the spin; higher twist quadrilinear operators can be formed through the action of \( \square \equiv \partial^\mu\partial_\mu \) on these operators.
The quadrilinear operators are generally highly degenerate: there are many different primary operators with the same twist and spin.
Where this happens we report the weighted average of CFT data that occurs in the crossing symmetry equation.
If the degenerate operators of a fixed twist \( \tau_0\) and spin \(l \) are labelled by an index \( i\), and \( a\up{0}_{\tau_0,l,i}\) are their OPE coefficients in the free theory, then, as per equation \eqref{eq:degavdef}, our results apply to the weighted average defined as follows:
\begin{equation} \label{eq:degavdefrep}
\degav{f_{\tau_0,l}} \equiv \frac{\sum_i a_{\tau_0,l,i}\up{0} f_{\tau_0,l,i}}{\sum_i a_{\tau_0,l,i}\up{0} } \,.
\end{equation}

The subscripts on our results indicate the \( U(N_f)\) representation, twist and spin of the operators; e.g. \( \gamma_{S,2,6}\up{1} \) refers to the order \( \epsilon^1\) part of the anomalous dimension of the singlet operator of twist 2 and spin 6.
Furthermore, we give results for the \( \alphahat \), which are related to the multiplicative OPE coefficient corrections \( \alpha \) by
\begin{equation} \label{eq:ahatdefRep}
 \widehat{\alpha}_{\tau_0,l}\up{1} = \alpha_{\tau_0,l}\up{1} - \frac{1}{2 a_{\tau_0,l}\up{0}}\partial_{l} \left(a_{\tau_0,l}\up{0}\gamma_{\tau_0,l}\up{1} \right) = \alpha_{\tau_0,l}\up{1} - \frac{1}{2} \partial_{l}  \gamma_{\tau_0,l}\up{1} - \frac{1}{2}\gamma_{\tau_0,l}\up{1} \partial_{l} \log a_{\tau_0,l}\up{0} \, .
\end{equation}

\subsection*{The \(d = 2+\epsilon \) expansion}
For the singlet sector we find a non-trivial solution at first order in \( \epsilon \) that depends on three constants: the external operator dimension \( \gamma\up{1}_\cO\), the central charge correction \( c_T\up{1} \), and a constant \(\beta\).

The bilinear currents satisfy
\begin{align}
\gamma_{S,0,l}\up{1} &= \beta\left( S_1(l-1) - 1 \right)\, ,\label{eq:2dBilAnomRep}\\
\ahat_{S,0,l}\up{1} &= \left( 2\gamma_\cO\up{1} +\beta\right) S_1(l-1) + \xihat_{-1}\, ,\label{eq:2dBilOPERep}
\end{align}
where \( \xihat_{-1} \) can be determined in terms of \( \gamma\up{1}_\cO\), \( c_T\up{1} \), and \(\beta\) through the equation for the stress-tensor OPE coefficient:
\begin{equation} \label{eq:centralChargeRelationApp}
a_{S,0,2} = \frac{1}{c_T}\frac{d^2}{(d-1)^2} \Delta_\cO^2\, .
\end{equation}
For the quadrilinear operators of twist \( \tau_0 = 2,4,6,\dots \), we find that
\begin{align} 
\degav{\gamma_{S,\tau_0 ,l}^{(1),inf.}} &= \beta_{\tau_0} S_1\left(l+\frac{\tau_0}{2}-1\right) +\kappa_{\tau_0}\, ,\label{eq:2dQuadAnomFormRep} \\
\degav{\ahat_{S,\tau_0,l}^{(1),inf.}} &= \alphahat_{\tau_0} S_1\left(l+\frac{\tau_0}{2}-1\right)+ \ksihat_{\tau_0}\, ,\label{eq:2dQuadOPEFormRep}
\end{align}
where
\begin{align}
\beta_{\tau_0} &= - \frac{\eta}{N+\eta} \beta \, , \label{eq:2dQuadBetaRep}\\
\kappa_{\tau_0} &= 2 \gamma_\cO\up{1} \frac{N+2\eta}{N+\eta} + \frac{\eta \beta}{N+\eta} \left(2-S_1\left(\frac{\tau_0}{2} -1 \right) + \frac{1}{2} \delta_{\tau_0,2}\right) \, ,\label{eq:2dQuadKappaRep}\\
\ahat_{\tau_0} &= 2\gamma_\cO\up{1} + \frac{\eta \beta}{N+\eta} \left(1- 2 S_1\left(\frac{\tau_0}{2}-1\right) + S_1\left(\tau_0-2\right) + \frac{1}{4}\delta_{\tau_0,2} \right)  \, , \label{eq:2dQuadAlphaRep}\\
\ksihat_{\tau_0} &= \frac{\eta}{N+\eta} \left( \xihat_{-1} + \beta \ksihat_{\tau_0}\up{\beta} + \gamma_\cO\up{1} \ksihat_{\tau_0}\up{\gamma_\cO}  \right)  \, , \label{eq:2dQuadXiRep}
\end{align}
where we defined $\eta = (-1)^{\frac{\tau_0}{2}}$, and where, in equation \eqref{eq:2dQuadXiRep}:
\begin{align}
\ksihat_{\tau_0}\up{\beta} &= \zeta_2 + 3 S_1\left( \frac{\tau_0}{2}-1 \right)-S_1\left( \frac{\tau_0}{2}-1 \right)^2 - 2 S_1\left( \tau_0- 2 \right) \nonumber\\
& \qquad\qquad + S_1\left( \frac{\tau_0}{2}-1 \right)S_1\left( \tau_0- 2 \right) + \frac{1}{2} S_2 \left(\frac{\tau_0}{2}-1 \right) - \frac{5}{4}\delta_{\tau_0,2}\, , \label{eq:2dQuadXiBetaRep}  \\
\ksihat_{\tau_0}\up{\gamma_\cO} &= 6 S_1\left( \frac{\tau_0}{2}- 1 \right) - 4 S_1\left( \tau_0- 2 \right) - \delta_{\tau_0,2} + N\eta \left( 4 S_1\left( \frac{\tau_0}{2}- 1 \right) - 2 S_1\left( \tau_0- 2 \right) \right)\, ,\label{eq:2dQuadXiGammaRep}
\end{align}
where \(\zeta_2 = \zeta(2) = \frac{\pi^2}{6}\).

Furthermore, there is a finite support solution for the quadrilinears of twist \(\tau_0 = 2,4,6\dots\), taking the form
\begin{align}
\degav{\gamma^{(1),fin.}_{S,\tau_0,0}} = \frac{N}{N+\eta} \frac{1}{\tau_0 -1} \left(\gamma_{fin} + \frac{\beta}{4N}\left(1-\delta_{\tau_0,2}\right) \right)\, ,
\end{align}
where \(\gamma_{fin} \) is a constant not fixed by our analysis.

\subsubsection*{Gross-Neveu model}
The above solution reduces to the Gross-Neveu model when \( \beta =0 \), yielding:
\begin{align}
\gamma_{S,0,l}\up{1} &= 0\, ,\label{eq:2dGNBilAnomRep}\\
\alpha_{S,0,l}\up{1} &= 2\gamma_\cO\up{1}  S_1(l-1)\, .\label{eq:2dGNBilOPERep}
\end{align}
Furthermore the quadrilinear operators have corrections of the form 
\begin{align} 
\degav{\gamma_{S,\tau_0 ,l}\up{1}} &= \kappa_{\tau_0}\, ,\label{eq:2dGNQuadAnomFormRep} \\
\degav{\ahat_{S,\tau_0,l}\up{1}} &= 2\gamma_\cO\up{1} S_1\left(l+\frac{\tau_0}{2}-1\right)+ \ksihat_{\tau_0}\, ,\label{eq:2dGNQuadOPEFormRep}
\end{align}
where
\begin{align}
\kappa_{\tau_0} &= 2 \gamma_\cO\up{1} \frac{N+2\eta}{N+\eta} \, ,\label{eq:2dGNQuadKappaRep}\\
\ksihat_{\tau_0} &= \frac{\eta}{N+\eta} \gamma_\cO\up{1} \ksihat_{\tau_0}\up{\gamma_\cO}   \, , \label{eq:2dGNQuadXiRep}
\end{align}
with \( \eta \) and \( \ksihat_{\tau_0}\up{\gamma_\cO} \) as above.

From these results for the singlet operators in the Gross-Neveu model, we deduce results for the non-singlet bilinear operators. 
Specifically, we find for the bilinear adjoint operators of even spin \( l \geqslant 2\):
\begin{align}
\gamma^{(1)}_{A,0,l} &= 0 \, , \\
\lambda_{AAl}^{(1)} &= \gamma_{\cO^A}^{(1)} S_1(l-1) + k_A \, , \\
\lambda_{SAl}^{(1)} &= \frac{1}{2}\left( \gamma_\cO^{(1)}+\gamma_{\cO^A}^{(1)} \right) S_1(l-1) + k_{SA} \, .
\end{align}
Here the \(\lambda_{\bullet\bullet \bullet} \up{1}\) are the (multiplicative) corrections to the (non-squared) OPE coefficients \(  c_{\bullet\bullet \bullet} \), with \(c_{AAl} = c_{\cO^A\cO^A J_{S,l}} \) and \(c_{SAl} = c_{\cO\cO^A J_{A,l}} \).
Furthermore, we have found that for bilinear currents of odd spin $l$:
\begin{equation}
\gamma_{A,0,l}^{(1)} = 0 \, .
\end{equation}

Furthermore we find that the corrections to the bilinear anomalous dimensions, to second order in \( \epsilon\), are of the form
\begin{align}
\gamma_{S,0,l}\up{2} &=0\, ,  & l \geqslant 2 \text{ even,}\\
\gamma_{A,0,l}\up{2} &=\gamma_{A,0,2}\up{2} \, ,  & l \geqslant 2 \text{ even,}\\
\gamma_{A,0,l}\up{2} &=\gamma_{A,0,3}\up{2}\left( 1 - \delta_{1,l}\right) \, ,  & l \geqslant 1 \text{ odd.}
\end{align}

The bilinear anomalous dimensions match known results for the Gross-Neveu model in $2+\epsilon$ dimensions, found for example in \cite{Giombi:2017rhm}.

\subsection*{The \( d= 4-\epsilon \) expansion: the Gross-Neveu-Yukawa model}
Our results are for the first order anomalous dimensions of the bilinear currents.
For the adjoint bilinear currents, these are:
\begin{equation}
\gamma_{A,l}\up{1} = 2\gamma_\psi^{(1)}\left( 1 - \frac{2}{l(l+1)}\right)\, ,
\end{equation}
for both odd and even spin $l\geqslant 1$.

The singlet bilinear currents $J_{\psi,l} \sim \psibar \gamma\partial^{l-1} \psi$ mix with the currents $J_{\sigma,l} \sim \sigma \partial^l \sigma$, and the anomalous dimensions of the resulting primary operators are, for even spin $l \geqslant 2$,
\begin{align}
\gamma_{S,l}\up{1} = 2\gamma_\psi^{(1)}\left( \frac{N+1}{2} -\frac{1}{l(l+1)} \pm
\frac{ \sqrt{4 + (-4 + 20 N)l(l+1)+ (N-1)^2 l^2(l+1)^2 }}{ 2 l(l+1)} \right)  ,
\end{align}
which were found as the eigenvalues of the following matrix
\begin{equation}
H = 2\gamma_\psi^{(1)}
\begin{pmatrix}
 1- \frac{2}{l(l+1)} & \pm\frac{2\sqrt{N}}{\sqrt{l(l+1)}} \\
\pm\frac{2\sqrt{N}}{\sqrt{l(l+1)}} & N 
\end{pmatrix}.
\end{equation}
This reproduces the results in \cite{Giombi:2017rhm}.

\section{Boundary term} \label{app:boundaryTerm}
In this appendix we try to make precise the statement from section \ref{sec:CrossFurtherAnalysis} that 
\begin{equation}
\sum_{\tau_0,l} \partial_{l} \left( a_{\tau_0,l}\up{0} \gamma_{\tau_0,l}\up{1}G_{\tau,l}(u,v)  \right)
\end{equation}
is a `boundary term' that does not contain any enhanced divergences in \( v \).

To this end, consider a function \( f: \mathbb{R}\rightarrow \mathbb{R} \), arising from some function \( \tilde{f}: \mathbb{N}\rightarrow \mathbb{R} \) that has been suitably analytically continued to have some desirable behaviour at infinity, and decays suitably quickly at infinity.
We show that under some reasonable assumptions, \( \sum_l f'(l) \) is a boundary term.

Fix an \( N \in \mathbb{N} \) and consider 
\begin{equation}
 f(N) - f(0) = \sum_{l=0}^{N-1} [f(l+1)-f(l)].
\end{equation}
By the intermediate value theorem, for each \(l\) there exists a \( \xi_l \in (l,l+1) \) such that \( f(l+1)-f(l) = f'(\xi_l) \).
Therefore
\begin{equation}
 \lim_{N\rightarrow \infty} f(N) - f(0) = \sum_{l=0}^{\infty} f'(\xi_l)\, .
\end{equation}
The left-hand side is clearly a boundary term, so that we are done if we can relate the right-hand side to \( \sum_l f'(l) \).
This can be done for example if \( f'(l) \) is monotonic; in fact, since we are interested in enhanced divergences, we do not care about finite sums and may in fact only demand that \( f'(l) \) is monotonic for some \( l > L \), and from numerical explorations we indeed find that this holds for the sums \( \sum_{l} \partial_{l} \left( a_{\tau_0,l}\up{0} \gamma_{\tau_0,l}\up{1}G_{\tau,l}(u,v)  \right) \) encountered in this paper.

\bibliographystyle{JHEP}
\bibliography{Crossing}

\end{document}